\documentclass[%
 aip,jcp,
 amsmath,amssymb,
reprint%
]{revtex4-1}

\usepackage{graphicx}
\usepackage{dcolumn}
\usepackage{bm}

\usepackage[utf8]{inputenc}
\usepackage[T1]{fontenc}
\usepackage{mathptmx}
\usepackage{etoolbox}

\usepackage{caption}
\usepackage{subcaption}
\usepackage{xcolor}

\newcommand{\ct}[1]{\begin{tabular}{l} #1 \end{tabular}}

\makeatletter
\def\@email#1#2{%
 \endgroup
 \patchcmd{\titleblock@produce}
  {\frontmatter@RRAPformat}
  {\frontmatter@RRAPformat{\produce@RRAP{*#1\href{mailto:#2}{#2}}}\frontmatter@RRAPformat}
  {}{}
}%
\makeatother
\begin{document}

\preprint{AIP/123-QED}

\title[SANN adapted for systems with low CN]{Solid-angle based nearest-neighbor algorithm adapted for systems with low coordination number}
\author{A. Ulug\" ol}
 \affiliation{%
Soft Condensed Matter and Biophysics Group, Debye Institute for Nanomaterials Science, Utrecht University, Princetonplein 1, Utrecht, 3584 CC, Netherlands
}%
\email{a.ulugol@uu.nl.}

\author{F. Smallenburg}%
\affiliation{Laboratoire de Physique des Solides, Université Paris-Saclay, Orsay, 91405, France}%

\author{L. Filion}
\affiliation{%
Soft Condensed Matter and Biophysics Group, Debye Institute for Nanomaterials Science, Utrecht University, Princetonplein 1, Utrecht, 3584 CC, Netherlands
}%

\date{\today}

\begin{abstract}
Nearest-neighbor identification is central to the analysis of local structure in condensed matter systems. The solid-angle-based nearest-neighbor (SANN) algorithm is widely used offering a parameter-free and computationally efficient alternative to cutoff- or Voronoi-based methods. Unfortunately, however, in systems with low coordination numbers, SANN tends to identify many particles as neighbors that are outside the nearest neighbor shell. Here, we propose a solution to this problem.  Specifically, we propose a geometric modification, the ``inscribed circle modification'', that resolves systematic overcounting in low-coordination lattices without introducing free parameters. We benchmark the modified algorithm (mSANN) against Voronoi and the original SANN algorithm in crystalline, quasicrystalline, and heterogeneous systems, and demonstrate that it provides robust and low-cost neighbor identification across both two and three dimensions.
\end{abstract}

\maketitle

\section{Introduction}\label{sec1}

Identifying nearest neighbors (NNs) is central to the analysis of local structure in many-body systems, and underlies a wide range of order parameters and classification methods.\cite{steinhardt1983bond,lechner2008accurate,malins2013identification,reinhart2017machine,boattini2019unsupervised} Despite its prevalence, no unique definition of NNs exists, and several algorithms are routinely employed.

The most direct approach is to impose a fixed-distance cutoff. A typical choice is the first minimum of the pair correlation function $g(r)$, which captures the first coordination shell. While effective in homogeneous systems, this definition becomes ambiguous in the presence of density gradients or phase coexistence, where a single cutoff cannot account for spatial variations.

An alternative is the Voronoi construction, which partitions space into polyhedra that tessellate the system.\cite{fortune1995voronoi} Two particles are defined as neighbors if their Voronoi cells share a face in 3D or edge in 2D. This approach is geometrical and parameter-free, and is often applied to inhomogeneous systems. However, it suffers from two drawbacks: sensitivity to thermal fluctuations, which leads to fluctuations in the number of neighbors even in a stable lattice, and systematic overcounting in low-coordination structures. Several weighted variants have been proposed,\cite{meijering1953interface,brostow1978construction,medvedev1986algorithm,malins2013identification} but these introduce additional parameters. Moreover, Voronoi tessellations are computationally demanding in three dimensions, which limits their use in on-the-fly analyses.

In 2012, van Meel \textit{et al.} introduced the solid-angle-based nearest-neighbor (SANN) algorithm as a parameter-free and computationally inexpensive alternative.\cite{van2012parameter} SANN assigns a solid angle to each candidate neighbor and defines the cutoff radius such that the sum of solid angles completes a sphere. This construction satisfies four desirable criteria: (i) absence of free parameters, (ii) robustness to thermal noise, (iii) low computational cost, and (iv) applicability to inhomogeneous systems. Nevertheless, the method has two limitations. First,  a direct extension, such as presented in Ref.~\onlinecite{isobe2016hard},  is less efficient in two dimensions as solving for the cutoff radius requires a numerical root-finding step. A solution to this issue was presented in Ref.~\onlinecite{mugita2024simple}, where they reformulate the problem as an existence-uniqueness problem (which they call SANNex).  Second, like Voronoi, it overestimates neighbor counts in low-coordination lattices (e.g. honeycomb or diamond). The original work\cite{van2012parameter} suggested introducing a tuning parameter to address the latter, at the expense of losing strict parameter-free character. Other extensions to SANN include an effort to take into consideration anisotropy in the environment \cite{staub2020parameter} and a variation aimed at making it more applicable for polydisperse systems (SANNR) \cite{varela2025solid}. 

In this paper, we revisit the foundations of SANN and examine in detail two modifications. First, following Ref. \onlinecite{mugita2024simple} we reformulate the cutoff condition to enable an efficient two-dimensional implementation, and then look in detail at this change. Second, we introduce a simple geometric modification, the ``inscribed circle (sphere) trick'', that eliminates systematic overcounting in low-coordination lattices without introducing free parameters. We compare the modified SANN (mSANN) with Voronoi and the original SANN across crystalline, quasicrystalline, and coexistence systems in both two and three dimensions, and show that it provides robust and efficient neighbor identification across a broad range of environments.

\section{Method description}

We begin by revisiting the original formulation of the solid-angle-based nearest-neighbor (SANN) algorithm \cite{van2012parameter} and then present its extension to two dimensions \cite{isobe2016hard, mugita2024simple}. The discussion proceeds in three steps: (i) the three-dimensional definition, (ii) its extension to two dimensions, and (iii) a new variation (mSANN) that addresses overcounting of distant neighbors.

\subsection{Description and Extension of SANN}\label{ssec:description}

The fixed-distance cutoff method defines neighbors of particle $i$ as those $j$ with $|\mathbf{r}_{ij}|<R$, where $R$ is a global parameter. While straightforward, this definition requires system-dependent tuning. SANN eliminates this parameter by assigning each particle $i$ a local cutoff $R_i$, determined such that the angular space around $i$ is exactly filled by its neighbors.

For a trial cutoff $R_i$, each candidate neighbor $j$ spans a solid angle $\theta_{ij}$. SANN defines $R_i$ by requiring that the sum of these angles completes the full surrounding sphere: $4\pi$ in three dimensions, $2\pi$ in two. In practice, only the distances $r_{i,j}=|\mathbf{r}_{i,j}|$ of the closest neighbors need to be considered.

In three dimensions the $m$-neighbor shell radius $R_i^{(m)}$ is obtained from the linear relation
\begin{align}\label{eq:sann3d}
   4 \pi = \sum_{j=1}^m 2 \pi \left(1 - \frac{r_{i,j}}{R_i^{(m)}} \right),
\end{align}
which can be solved directly for $R_i^{(m)}$.

\begin{figure}[h]
\includegraphics[width=0.6\columnwidth]{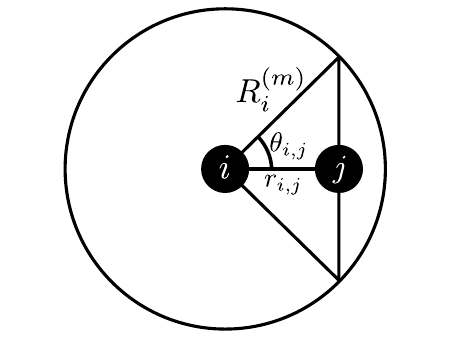}
\caption{Angle $\theta_{ij}$ associated with the chord passing through neighbor $j$. $R^{(m)}_i$ denotes the shell radius of particle $i$ with $m$ neighbors, and $r_{i,j}$ is the distance between $i$ and $j$.}
\label{fig:thetadef}
\end{figure}

In two dimensions, replacing the sphere by a circle yields the nonlinear condition
\begin{align}\label{eq:sann}
   2\pi =2\sum_{j=1}^m\theta_{ij}= 2\sum_{j=1}^m \arccos\!\left(\frac{r_{i,j}}{R_i^{(m)}}\right),
\end{align}
where $\theta_{ij}$ is half the central angle of the chord associated with neighbor $j$ as shown in Fig.~\ref{fig:thetadef}. Unlike Eq.~\eqref{eq:sann3d}, Eq.~\eqref{eq:sann} cannot be solved in closed form. It can be solved via a root-finding algorithm, at extra numerical cost.

To retain the efficiency of SANN in two dimensions, as shown in Ref. \onlinecite{mugita2024simple}, the problem can be cast in terms of existence rather than an explicit solution. Define
\begin{equation}
f_i(x;m) = -\pi + \sum_{j=1}^m \arccos \!\left(\frac{r_{i,j}}{x}\right), \quad x \geq r_{i,m}.
\end{equation}
Note that $f_i(R_i^{(m)};m)=0$ by construction. Since $\arccos$ is strictly decreasing, $f_i(x;m)$ is strictly increasing in $x$ and thus admits a unique root. It follows that $R_i^{(m)}$ lies within the interval $[r_{i,m},r_{i,m+1})$ if and only if
\begin{equation}
f_i(r_{i,m+1};m) > 0 \quad\mathrm{and}\quad f_i(r_{i,m};m-1) \leq 0.
\end{equation}

This observation leads to a simple iterative procedure: starting from $m=3$ (the minimum possible SANN coordination), increment $m$ until the above condition is satisfied. The corresponding $m$ then defines the coordination number of particle $i$, and the neighbor set is $\mathcal{N}_b(i)=\{j\}_{j=1}^m$. Unlike the original formulation, this approach generalizes directly to arbitrary dimension while avoiding nonlinear root finding.

\subsection{Algorithm}\label{ssec:algo}
A step-by-step procedure (analogous to Ref. \onlinecite{mugita2024simple}) to identify the nearest neighbors of particle $i$ as we have introduced in subsection \ref{ssec:description} is given by:
\begin{enumerate}
    \item Compute distances $r_{i,j}$ to all potential neighbors $\{j\}$ from particle $i$.
    \item Sort possible neighbors $\{j\}$ by their distance $r_{i,j}$ in increasing order such that $r_{i,j}\leq r_{i\ j+1}$.
    \item Set $m = 3$ (i.e., the minimum number of neighbors).
    \item Compute $f_i(r_{i\,m+1} ; m)$.
    \item if $f_i(r_{i\,m+1} ;m) \leq 0$ then increment $m$ by 1 and go back to step 4.
    \item Otherwise, $m$ is the number of neighbors for particle $i$, and $\mathcal{N}_b(i) = \{j\}_{j=1}^m$ is the associated set of neighbors.
\end{enumerate}

\subsection{Algorithm Properties}\label{ssec:algoprop}

The two-dimensional extension of SANN retains several properties of the original three-dimensional algorithm, while introducing a few modifications. In what follows we examine six key aspects:

\begin{enumerate}
    \item \textbf{Convergence:} the iterative procedure always terminates provided that a sufficient number of candidate neighbors exist.
    \item \textbf{Equidistant neighbors:} if multiple particles are equidistant from a given particle, then either all or none are identified as neighbors.
    \item \textbf{Pairwise symmetry:} neighbor relations need not be symmetric, though simple symmetrization schemes can be applied when required.
    \item \textbf{Tiling generation:} in two dimensions, neighbor relations can be used to construct a consistent tiling of the plane.
    \item \textbf{Local area estimation:} the algorithm provides an approximate measure of the area associated with each particle.
    \item \textbf{Higher-order neighbors:} the identification can be iterated to obtain next-nearest neighbors and beyond.
\end{enumerate}

Properties (3) and (6) are unchanged from the original three-dimensional formulation \cite{van2012parameter}. Properties (1), (2), and (5) require modified arguments in two dimensions, and property (4) is specific to the two-dimensional extension introduced here.
\subsubsection{Convergence}

\textbf{Lemma.} The iterative procedure defining $R_i^{(m)}$ converges, provided a sufficient number of candidate neighbors exist.

\textbf{Proof.}  
Let $R_i^{(m)}$ and $R_i^{(m+1)}$ denote the shell radii obtained when $m$ and $m+1$ neighbors are included. By definition,
\[
f_i(R_i^{(m)};m)=0, 
\quad 
f_i(R_i^{(m+1)};m+1)=0,
\]
where $f_i(x;m)=-\pi+\sum_{j=1}^m \arccos(r_{i,j}/x)$. Taking the difference gives
\begin{align}\label{eq:convtrivial}
0 =& f_i(R_i^{(m+1)};m+1) - f_i(R_i^{(m)};m) \nonumber \\
  =& \arccos\!\left(\frac{r_{i,m+1}}{R_i^{(m+1)}}\right) 
   \nonumber\\&+ \sum_{j=1}^m 
     \Bigg[\arccos\!\left(\frac{r_{i,j}}{R_i^{(m+1)}}\right) 
   - \arccos\!\left(\frac{r_{i,j}}{R_i^{(m)}}\right)\Bigg].
\end{align}
The first term is non-negative since $r_{i,m+1} \leq R_i^{(m+1)}$. Hence the sum in Eq.~\eqref{eq:convtrivial} must be negative:
\[
\sum_{j=1}^m 
\left[\arccos\!\left(\frac{r_{i,j}}{R_i^{(m+1)}}\right) 
- \arccos\!\left(\frac{r_{i,j}}{R_i^{(m)}}\right)\right] \leq 0.
\]
Because $\arccos(x)$ is strictly decreasing, each term satisfies
\[
\arccos\!\left(\frac{r_{i,j}}{R_i^{(m+1)}}\right) 
\leq \arccos\!\left(\frac{r_{i,j}}{R_i^{(m)}}\right),
\]
implying
\[
R_i^{(m+1)} \leq R_i^{(m)}.
\]
Thus, the cutoff radius decreases monotonically with $m$, ensuring convergence.

\subsubsection{Equidistant neighbors}\label{sssec:equidist}

\textbf{Lemma.}  
If two or more particles are equidistant from particle $i$, then either all or none of them are identified as neighbors.

\textbf{Proof.}  
Suppose, for contradiction, that two particles satisfy $r_{i,m}=r_{i,m+1}$, yet only one is included, so that the algorithm converges with $\mathcal{N}_b(i)=m$. This requires $f_i(r_{i,m+1};m)>0$. Since $r_{i,m}=r_{i,m+1}$, it follows that $f_i(r_{i,m};m)>0$ as well.  

Now observe that
\begin{align}
f_i(r_{i,m};m) 
 &= -\pi + \sum_{j=1}^m \arccos\!\left(\frac{r_{i,j}}{r_{i,m}}\right) \nonumber \\
 &= -\pi + \sum_{j=1}^{m-1} \arccos\!\left(\frac{r_{i,j}}{r_{i,m}}\right) \nonumber \\
 &= f_i(r_{i,m};m-1).
\end{align}
Hence $f_i(r_{i,m};m-1)>0$, which would force the algorithm to converge with $\mathcal{N}_b(i)=m-1$. This contradicts the assumption that exactly one of the equidistant neighbors was included.  

Therefore, the algorithm cannot separate equidistant particles: they are either all included or all excluded. 

\subsubsection{Pairwise symmetry}\label{sssec:pairwisesym}

\textbf{Property.}  
SANN does not guarantee pairwise symmetry: if particle $j$ is identified as a neighbor of $i$, the converse need not hold.

\textbf{Explanation.}  
This asymmetry arises because SANN assigns a distinct cutoff radius $R_i$ to each particle. Thus $j$ may lie within $R_i$ while $i$ lies outside $R_j$. Such cases are rare and typically occur only in systems with strong density gradients. For most applications, this lack of symmetry does not affect neighbor-dependent observables.

\textbf{Symmetrisation strategies.}  
When pairwise symmetry is required, two post-processing options are available, as proposed by the original work\cite{van2012parameter}:
\begin{itemize}
    \item \emph{Additive symmetrisation:} if $(i,j)$ is asymmetric, add the missing link so that $i$ and $j$ become mutual neighbors.
    \item \emph{Subtractive symmetrisation:} if $(i,j)$ is asymmetric, remove the link entirely so that neither particle lists the other as neighbor.
\end{itemize}
The choice between these schemes is application-dependent, as neither is inherently preferable.

\subsubsection{Tiling generation}\label{sssec:tiling}

In two-dimensional systems it is often desirable to construct a tiling representation of the particle network, for example when analyzing quasicrystalline order. The 2D extension of SANN provides a natural basis for such a construction.

\textbf{Procedure.}  
A tiling can be generated in four steps:
\begin{enumerate}
    \item Identify the nearest neighbors of all particles using SANN.
    \item Apply a symmetrisation scheme (Sec.~\ref{sssec:pairwisesym}) to ensure pairwise consistency.
    \item Connect each pair of neighboring particles with a line segment.
    \item If two line segments intersect, remove the longer one.
\end{enumerate}

The resulting network defines a tiling in which each particle corresponds to a vertex. This approach is straightforward to implement and yields tessellations consistent with the local connectivity identified by SANN.

\subsubsection{Local area}

Estimating local particle areas can be useful, for instance when analyzing phase coexistence. Voronoi tessellation remains the natural choice for such tasks, since it provides exact local cells. Nevertheless, SANN can serve as a lightweight alternative when a full Voronoi construction is unnecessary or computationally costly.

In three dimensions, the cutoff radius $R_i$ follows directly from the SANN condition and defines a sphere of radius $R_i$, yielding a local volume estimate~\cite{van2012parameter}. In two dimensions, our extension determines only the interval in which $R_i$ must lie. From this point, two simple options are available: (i) solving for $R_i$ explicitly using a root-finding procedure, or (ii) approximating $R_i$ by the midpoint of the interval. The local area is then assigned as the disk of radius $R_i$.

A more geometrically faithful SANN-based definition of local areas can, in principle, be obtained by exploiting the tiling construction described in Sec.~\ref{sssec:tiling}. After symmetrizing the SANN neighbor relations and constructing the planar tiling by connecting neighboring particles, one may form the \emph{dual tiling}. The dual tiling is obtained by placing a vertex at the centroid of each face of the original tiling and connecting centroids of adjacent faces by edges that cross the original edges. The resulting dual faces define polygonal cells surrounding each particle.

The area of each such dual cell can then be interpreted as the \emph{SANN cell} associated with that particle, providing a local area measure that strictly respects the SANN neighbor topology by construction. Unlike the disk-based approximation, this approach yields polygonal cells whose number of edges exactly matches the SANN coordination number.

We emphasize, however, that this construction relies on additional geometric assumptions. For instance, the SANN-based tiling must be simply-connected for a well-defined dual to exist, and the faces of the primal tiling should be convex for the centroid-based construction to be unambiguous. These conditions are typically satisfied in ordered and weakly disordered two-dimensional systems, including the crystalline and quasicrystalline examples considered in this work. In heterogeneous environments, such as regions with large density gradients or highly irregular local connectivity, the tiling may become non-convex or poorly conditioned, in which case the dual construction can fail or produce ill-defined cells. In such situations, the disk-based area estimate or Voronoi tessellation may be more appropriate.

Although this dual-tiling approach introduces additional assumptions and is more involved than assigning a disk area; when the conditions are met, it provides a geometrically well-defined SANN-based cell construction whose topology strictly reflects the neighbor relations identified by the algorithm. We do not pursue this construction further here, but note that it offers a natural route to defining local areas or free volumes that are consistent with SANN neighbor identification. Related SANN-based cell and free-volume constructions are discussed in Ref.~\onlinecite{mugita2024simple}.

\subsubsection{Next-nearest neighbors}

The identification of next-nearest neighbors using SANN has been discussed in the original three-dimensional formulation by Ref.~\onlinecite{van2012parameter}. There, an iterative procedure was proposed: (i) determine the nearest neighbors of a particle using the algorithm in Sec.~\ref{ssec:algo};  
(ii) remove these neighbors from the candidate set;  
(iii) rerun the algorithm beginning from step 3. 

The new set of neighbors corresponds to the next-nearest neighbors of the particle. Repeating this procedure allows identification of the $n^\text{th}$ neighbor shell in a consistent manner. While this approach is well defined, its accuracy was shown to be limited.\cite{van2012parameter}

Specifically, Ref.~\onlinecite{van2012parameter} reported that for non-crystalline systems, this procedure tends to truncate the neighbor shell before the second minimum of the radial distribution function $g(r)$, whereas for crystalline systems it typically extends beyond the second minimum, including particles from higher shells \cite{van2012parameter}.

More recently, Ref.~\onlinecite{mugita2024simple} proposed a modified version of the 2D SANN algorithm for higher-order neighbors, in which the angular completion condition, Eq.\ref{eq:sann}, is modified by replacing the total angle $2\pi$ with $2\pi n$ for the identification of the $n^\text{th}$ neighbor shell. For the second shell in 2D disordered bidisperse systems, this modification was shown to yield reasonable agreement with the position of the second minimum of $g(r)$.~\cite{mugita2024simple}

We note, however, that this modification is empirical in nature and does not resolve the fundamental limitation identified in Ref.~\onlinecite{van2012parameter}. In particular, for crystalline systems, multiplying the angular target by $n$ is expected to push the effective cutoff even further beyond the relevant minimum of $g(r)$, thereby intensifying the overextension into higher neighbor shells. 

\subsection{Geometric modification based on the inscribed circle/sphere (mSANN)}

Up to this point, we have emphasized the advantages of parameter-free neighbor definitions such as SANN and Voronoi over fixed-distance cutoffs. However, both SANN and Voronoi systematically overcount neighbors in lattices with low coordination number (CN) \cite{van2012parameter}. Here, we define ``overcounting" as the identification of particles that lie beyond the first minimum of the radial distribution function as neighbors -- in the absence of density gradients. As we will see later in this paper, this includes e.g. the honeycomb (CN=3) and square (CN=4) lattices in two dimensions, and the $\alpha$-graphite (CN=3), diamond (CN=4), simple cubic (CN=6), and body-centered cubic (CN=8) lattices in three dimensions. While in principle there is no ``correct'' definition of nearest neighbors, this often leads to the inclusion of neighbors that one would not intuitively consider to be part of the first neighbor shell.

For Voronoi-based algorithms, a common remedy for this issue is to weight neighbors by the area (or length) of their shared boundary,\cite{meijering1953interface,brostow1978construction,medvedev1986algorithm,malins2013identification} but this introduces additional parameters.  
For SANN, van Meel \textit{et al.} \cite{van2012parameter} proposed modifying Eq.~\eqref{eq:sann3d} by multiplying $4\pi$ with a free parameter. While this alleviates overcounting, it breaks the strict parameter-free character of the algorithm. Here we propose an alternative, parameter-free correction based on the underlying geometry.

\subsubsection*{Geometric interpretation of SANN in two dimensions}

For a particle $i$ and a trial neighbor count $m$, the two-dimensional SANN construction determines a radius $R_i^{(m)}$ such that the sum of the central angles associated with the $m$ closest neighbor candidates equals $2\pi$, as expressed by Eq.~\eqref{eq:sann}. Concretely, for each candidate neighbor $j$, one considers the chord of a circle of radius $R_i^{(m)}$ whose midpoint coincides with the position of particle $j$; the corresponding central angle is uniquely determined by the distance $r_{i,j}$.

\begin{figure}
\includegraphics[width=\columnwidth]{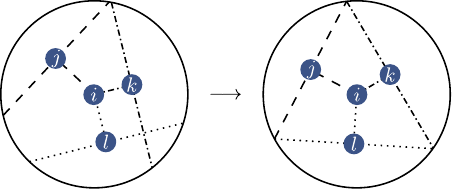}
\caption{Interpretation of Eq.~\eqref{eq:sann} in two dimensions. Local rotations around particle $i$ map neighbor positions onto the midpoints of a cyclic polygon, with $R_i$ as the radius of the circumscribed circle.}
\label{fig:howsannworks}
\end{figure}

For a circle of fixed radius, a chord is uniquely specified (up to a rotation about the center) by its central angle. Consequently, any collection of $m$ chords whose central angles sum to $2\pi$ can be rearranged by rotation to form a closed cyclic $m$-gon with the same circumscribed circle of radius $R_i^{(m)}$ (Fig.~\ref{fig:howsannworks}). This rearrangement is purely geometric and does not assume any specific angular ordering of the neighbors. It relies only on the neighbor distances $\{r_{i,j}\}$, consistent with the distance-based nature of SANN.

This construction holds for arbitrary neighbor configurations. Angular order plays no role in the SANN criterion and is therefore irrelevant for the existence of the associated cyclic polygon.

\subsubsection*{Ideal lattices and overcounting}

A particularly transparent illustration of SANN's overcounting behavior arises in the special case where all $m$ neighbor candidates are equidistant from the central particle. In this case, all associated chords subtend identical central angles, and the rearranged cyclic polygon becomes a regular $m$-gon. This situation is realized, for example, in ideal crystalline lattices, which additionally impose angular order. While this angular order is not required for the following argument, ideal lattices provide a convenient reference case.

For a regular $m$-gon, the circumscribed circle has radius $r_\mathrm{cir}$, which coincides with the SANN radius $R_i^{(m)}$, while the inscribed circle has radius $r_\mathrm{in}$. The neighbor positions lie on the inscribed circle. For low coordination numbers $m$, the ratio $r_\mathrm{in}/r_\mathrm{cir}$ is small, meaning that the SANN radius significantly exceeds the actual nearest-neighbor distance. As a result, SANN is likely to enclose not only the true nearest neighbors but also particles from higher coordination shells.

\begin{figure}
\includegraphics[width=\columnwidth]{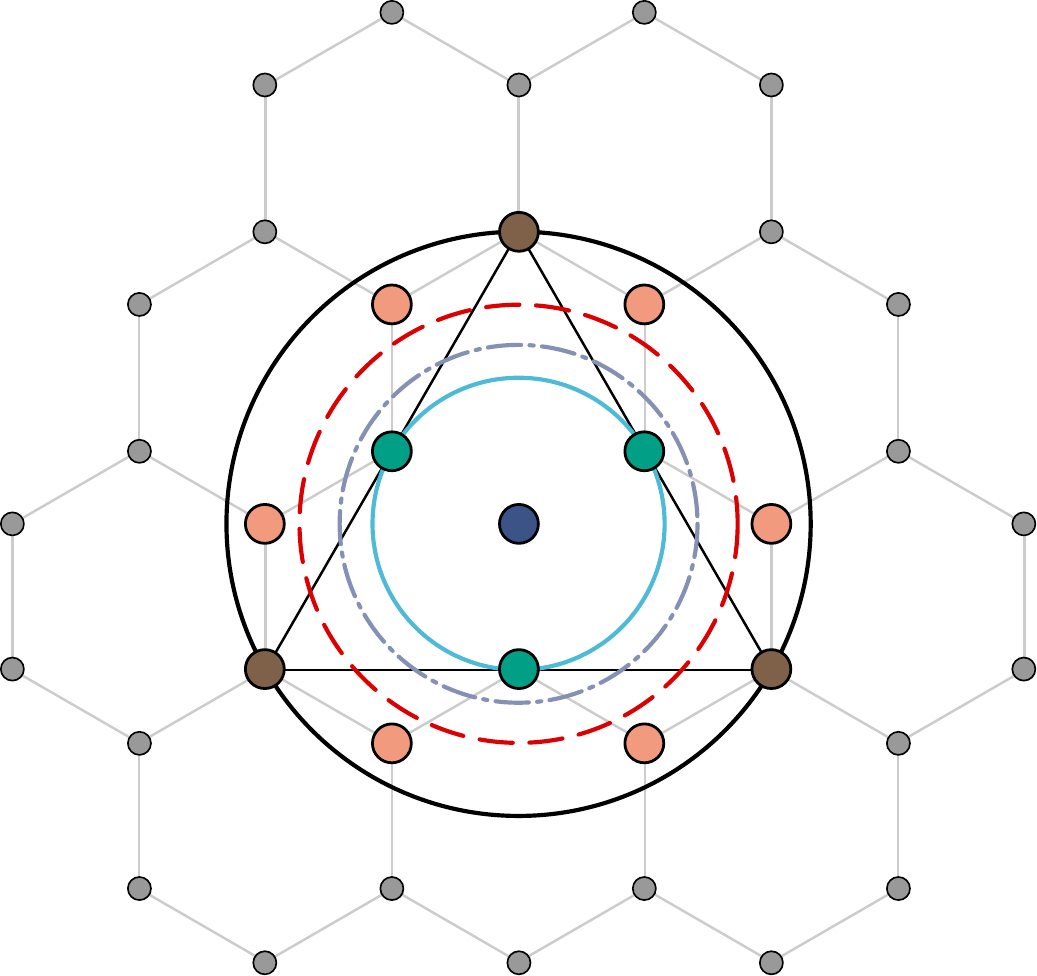}
\caption{Illustration of the first iteration ($m=3$) of the SANN construction applied to an ideal honeycomb lattice. The three nearest neighbors define a triangle whose side midpoints coincide with the neighbor positions. Solving Eq.~(\ref{eq:sann}) yields the black circle, which corresponds to the circumscribed circle of this triangle. Because this radius encloses next- and next-next-nearest neighbors of the honeycomb lattice, the SANN algorithm does not converge at m=3, leading to overcounting. The inscribed circle of the same triangle (green) would instead yield immediate convergence and the correct coordination number, but leaves no margin for thermal fluctuations. Also shown are circles with radii given by the arithmetic (red dashed) and geometric (gray dot-dashed) means of the inscribed and circumscribed radii. Both provide sufficient tolerance for fluctuations while excluding farther neighbors, resulting in correct convergence.}
\label{fig:honey}
\end{figure}

For example, in a perfect honeycomb lattice ($m=3$), the circumscribed circle of the associated triangle has radius $2a$ (with $a$ the lattice constant), which encloses both next- and next-next-nearest neighbors (Fig.~\ref{fig:honey}). This prevents the algorithm from converging to the correct coordination number. While a small value of $r_\mathrm{in}/r_\mathrm{cir}$ does not guarantee overcounting in general, it makes such overcounting highly likely in low-coordination environments.

\begin{figure*}
\includegraphics[width=\textwidth]{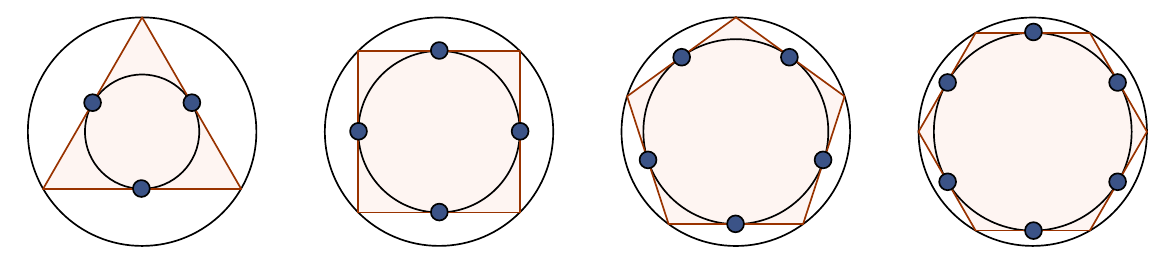}
\caption{Inscribed and circumscribed circles of the first four regular polygons.}
\label{fig:circles}
\end{figure*}

\subsubsection*{The inscribed circle modification}

The original SANN algorithm effectively uses the circumscribed circle of the associated cyclic polygon to define the neighbor shell. This choice makes SANN robust to thermal fluctuations, as small variations in neighbor distances are unlikely to remove a neighbor from the shell. However, in low-coordination environments this criterion can become overly permissive, leading to systematic overcounting.

At the opposite extreme, the most restrictive nearest-neighbor criterion would require all neighbors to be equidistant from the central particle. In the geometric construction above, this corresponds to using the inscribed circle of the regular $m$-gon, for which the neighbors lie exactly on the boundary. While such a criterion would eliminate overcounting in ideal lattices, it leaves no tolerance for thermal fluctuations and is therefore impractical in real systems.

The inscribed-circle modification introduced here should be understood as an interpolation between these two limiting cases. For a regular $m$-gon, the ratio of inscribed to circumscribed radius is
\begin{equation}\label{eq:inoutcircle}
    \frac{r_\mathrm{in}}{r_\mathrm{cir}} = \cos\!\left(\frac{\pi}{m}\right),
\end{equation}
where $r_\mathrm{in}$ and $r_\mathrm{cir}$ denote the radii of the inscribed and circumscribed circles, respectively, and $r_\mathrm{cir}$ coincides with the SANN radius $R_i^{(m)}$. The corresponding circles of the first four regular polygons can be seen in Fig.~\ref{fig:circles}.

Rather than replacing the SANN radius by $r_\mathrm{in}$, which would implicitly enforce equidistance, we shall introduce an interpolated radius $
\mathcal{R}_i^{(m)}$ between $r_\mathrm{in}$ and $r_\mathrm{cir}$. To achieve this, consider the interpolating function

\begin{equation}
k : \mathbb{R}^2 \to \mathbb{R}, \qquad (x,y) \mapsto k(x,y),
\end{equation}
which satisfies the following properties:
\begin{align}
\min(x,y) &\leq k(x,y) \leq \max(x,y), \label{p:bound}\\
k(\lambda x, \lambda y) &= \lambda\, k(x,y) \qquad \forall\, \lambda \in \mathbb{R}, \label{p:homo}\\
x_i > x_j &\;\Rightarrow\; k(x_i,y) > k(x_j,y),\label{p:mono} \\
y_i > y_j &\;\Rightarrow\; k(x,y_i) > k(x,y_j)\nonumber.
\end{align}
Without specifying the functional form of $k(x,y)$ explicitly, we define the interpolating radius

\begin{equation}
\begin{aligned}
        \mathcal{R}_i^{(m)} &= k\left(R_i^{(m)} , \frac{r_\mathrm{in}}{r_\mathrm{cir}} R_i^{(m)}\right),
        \\&=k\big(1,\cos(\pi/m)\big)R_i^{(m)},
\end{aligned}
\end{equation}
where we used the property \ref{p:homo} in the second line. For brevity, we introduce the shorthand notation
\begin{equation}
    k_m:= k\big(1,\cos(\pi/m)\big).
\end{equation}
Using the properties of $k(x,y)$, we find that $k_m$ satisfies:
\begin{align}
    \cos(\pi/m) \leq k_m \leq 1,\\
    m>n\Rightarrow k_m>k_n.
\end{align}
This choice corresponds to a generalized average of the inscribed and circumscribed radii in the ideal reference case. It does not assume equidistant neighbors in general, but instead, since $k_m\rightarrow1$ as $m\rightarrow\infty$, reduces to the original SANN radius for large $m$ where SANN is known to perform well. 

We refer to this parameter-free modification as the \emph{inscribed circle trick}, and the variation including this trick as modified SANN (mSANN). Operationally, this modification leads to a revised convergence criterion: For a given trial neighbor count $m$, convergence is achieved if the interpolated radius $\mathcal{R}_i^{(m)}$ lies within the interval $[r_{i,m},r_{i,m+1})$, which is equivalent to
\begin{equation}
f_i(r_{i,m+1}/k_m;m) > 0 \quad\mathrm{and}\quad f_i(r_{i,m}/k_{m-1};m-1) \leq 0.
\end{equation}

Thus, the practical implementation of mSANN proceeds exactly as in standard SANN when one replaces $f_i(r_{i\,m+1} ; m)$ with $f_i(r_{i\,m+1}/k_m ; m)$ in steps 4 and 5 of the algorithm.

However, we need to make sure this modified criterion converges:

\textbf{Lemma.} The modified iterative procedure defining $\mathcal{R}_i^{(m)}$ converges, provided a sufficient number of candidate neighbors exist.

\textbf{Proof.}  
As we have proved that SANN converges. It is enough to show that the convergence of SANN implies the convergence of modified SANN.

Suppose that for a particle $i$, SANN converges at $m_0$ nearest neighbors. This implies that
\[
0<f_i(r_{m_0+1};m_0)
\]
since $f_i(x;m)$ is monotonic increasing function and $0<k_m\leq1$, the following is also true:
\[
0<f_i(r_{m_0+1};m_0)\leq f_i(r_{m_0+1}/k_;m_0)
\]
so the mSANN convergence condition is also satisfied at $m_0.$ Therefore mSANN terminates no later than $m_0$. Therefore, mSANN converges irrespective of the choice of $k_m$.

For the remaining of the manuscript, we choose the arithmetic mean for $k(x,y)$ in 2D such that

\begin{equation}
k_m = \frac{1}{2}\big[1+\cos(\pi/m)\big].
\end{equation}

We note that this is an empirical choice: it yields the smallest correction that eliminates overcounting in the ideal honeycomb lattice while retaining sufficient tolerance for thermal fluctuations. Other interpolations are in principle possible; however, as we will show later in the manuscript (see also Appendix \ref{ap:spread}), this choice performs robustly across a wide range of non-ideal systems.

Furthermore, this choice breaks the strict equidistant-neighbor property (see Appendix \ref{ap:example} for an example); care must therefore be taken to verify robustness in cases with degeneracies.

To recover the strict equidistant-neighbor property, one might modify the algorithm to increment $m$ by the number of next equidistant points in step $5$ of the algorithm. This calls for a free parameter, $\epsilon$, to define equidistance in float point arithmetic, i.e., point $j$ and point $l$ are equidistant to point $i$ if $|r_{i,j}-r_{i,l}|<\epsilon$. We avoid this modification to keep the algorithm parameter-free.

\subsubsection*{Extension to Three-dimensional case}

An analogous construction applies in three dimensions. For a fixed trial neighbor count $m$, the SANN shell radius admits the closed-form expression~\cite{van2012parameter}
\begin{equation}\label{eq:rsann3d}
    R^{(m)}_i = \frac{\sum_{j=1}^m r_{i,j}}{m-2}.
\end{equation}

To expose the geometric content of this expression, we again consider an idealized reference case. If the $m$ nearest neighbors are equidistant from particle $i$, with distance $r_{NN}$, Eq.~\eqref{eq:rsann3d} reduces to
\begin{equation}
    R^{(m)}_i = \frac{m}{m-2}\, r_{NN}.
\end{equation}
This relation defines the three-dimensional analogue of the inscribed-to-circumscribed radius ratio encountered in two dimensions,
\begin{equation}
    \frac{r_{NN} }{R^{(m)}_i} = 1 - \frac{2}{m}.
\end{equation}

As in two dimensions, standard SANN corresponds to using the larger, more permissive radius $R^{(m)}_i$, which provides robustness against thermal fluctuations but leads to systematic overcounting in low-coordination lattices. The opposite extreme would be to enforce the equidistant-neighbor limit $r_{NN}$, which is overly restrictive in non-ideal systems. The purpose of the modification is therefore not to assume equidistance, but to interpolate between these two limiting behaviors in a coordination-number-dependent and parameter-free manner.

We therefore introduce a modified radius $\mathcal{R}^{(m)}_i$ that interpolates between $r_{NN}$ and $R^{(m)}_i$. Among the possible interpolations, we choose the geometric mean,
\begin{equation}
    \mathcal{R}^{(m)}_i = \sqrt{r_{NN} R^{(m)}_i}
    = R^{(m)}_i \sqrt{1-\tfrac{2}{m}}.
\end{equation}
This choice yields the smallest reduction of the SANN radius that eliminates systematic overcounting in low-coordination reference lattices, most notably the ideal BCC lattice, while ensuring that the modified radius smoothly converges back to the original SANN radius at high coordination numbers, since $\mathcal{R}^{(m)}_i \to R^{(m)}_i$ as $m \to \infty$. Note that here we adopt the convention that ideal BCC lattice has coordination number $8$, in particular, $8$ nearest neighbors and $6$ next-nearest neighbors.

Finally, we would like to emphasize that the original and modified SANN ((m)SANN) construction requires $m\geq 3$; consequently, (m)SANN is not applicable to environments with fewer than three neighbors, where the underlying geometric construction is ill-defined in both 2D and 3D.

\section{Results}

\subsection{Bulk phases}

We first benchmark Voronoi, SANN, and mSANN in bulk crystalline systems without density gradients. In two dimensions, we consider honeycomb, square, and hexagonal lattices, which have coordination numbers (CN) of $3$, $4$, and $6$, respectively. The honeycomb and square lattices were generated from Monte Carlo simulations\cite{frenkel2023understanding} of binary mixtures of hard disks, retaining only the large species. Specifically, for the honeycomb lattice we used a system with a size ratio of 0.386 and a packing fraction of 0.797, where the holes between large particles were filled by seven small particles each. For the square lattice we used a size ratio of 0.45 and a packing fraction of 0.803, with each square hole between large particles containing one small particle. The hexagonal lattice was obtained from a Monte Carlo\cite{frenkel2023understanding} simulation of $512$ Lennard-Jones disks in the canonical ensemble at density $\rho\sigma^2=0.9$ and interaction strength $\beta\epsilon=2$. In three dimensions, we analyze simple cubic (SC), body-centered cubic (BCC), face-centered cubic (FCC), hexagonal close-packed (HCP), cubic diamond, and $\alpha$-graphite. Ideal configurations with $1000$–$1372$ particles were prepared and perturbed by uncorrelated Gaussian displacements of standard deviation $0.05a$, where $a$ is the lattice constant.

\begin{figure*}
\begin{ruledtabular}
\begin{tabular}{cccc}
 Lattice&Voronoi&SANN&mSANN \\ \hline
\ct{Honeycomb}&
\ct{\includegraphics[width=0.28\textwidth]{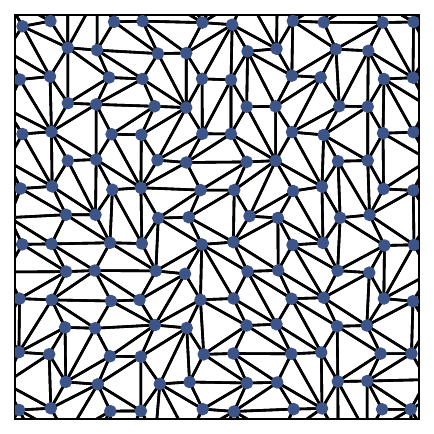}}&
\ct{\includegraphics[width=0.28\textwidth]{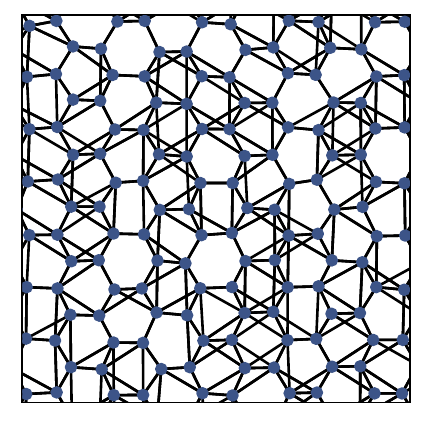}}&
\ct{\includegraphics[width=0.28\textwidth]{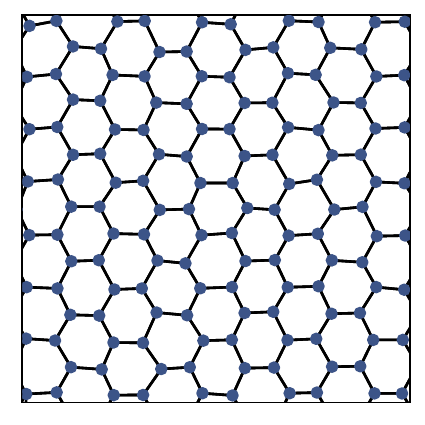}}
\\
\ct{Square}&
\ct{\includegraphics[width=0.28\textwidth]{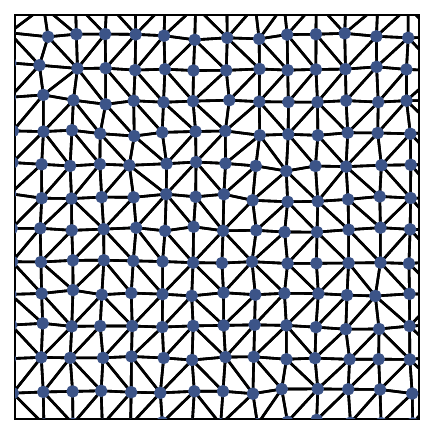}}&
\ct{\includegraphics[width=0.28\textwidth]{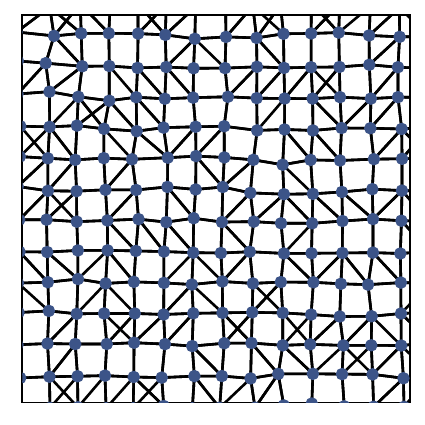}}&
\ct{\includegraphics[width=0.28\textwidth]{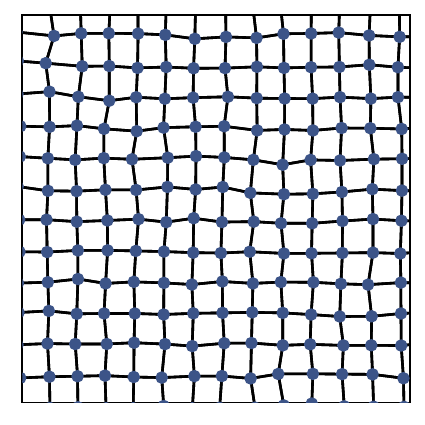}}
\\
\ct{Hexagonal}&
\ct{\includegraphics[width=0.28\textwidth]{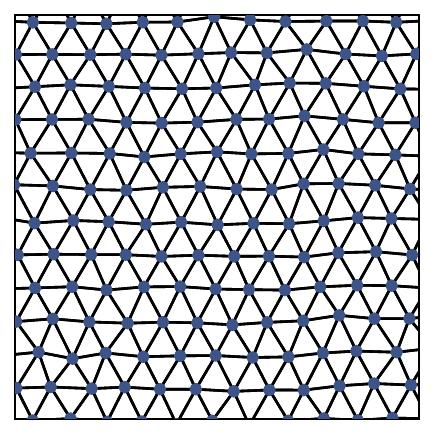}}&
\ct{\includegraphics[width=0.28\textwidth]{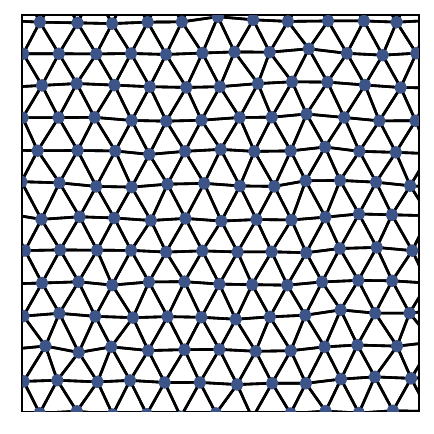}}&
\ct{\includegraphics[width=0.28\textwidth]{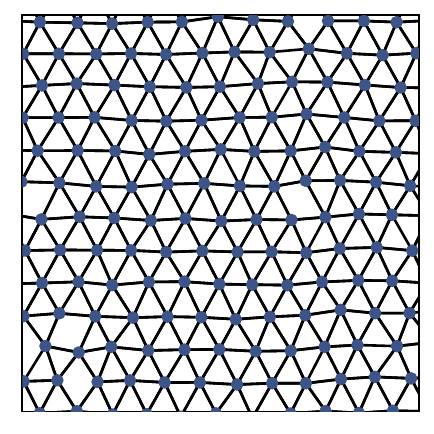}}
\\
\end{tabular}
\end{ruledtabular}
\caption{\label{fig:compare}Comparison of nearest-neighbor identification in 2D lattices using Voronoi, SANN, and mSANN. Dots represent particles and line segments connect the neighbors. Configurations include correlated thermal noise. Additive symmetrization is used for the visualization of SANN and mSANN.}
\end{figure*}

\begin{figure*}
\includegraphics[width=\textwidth]{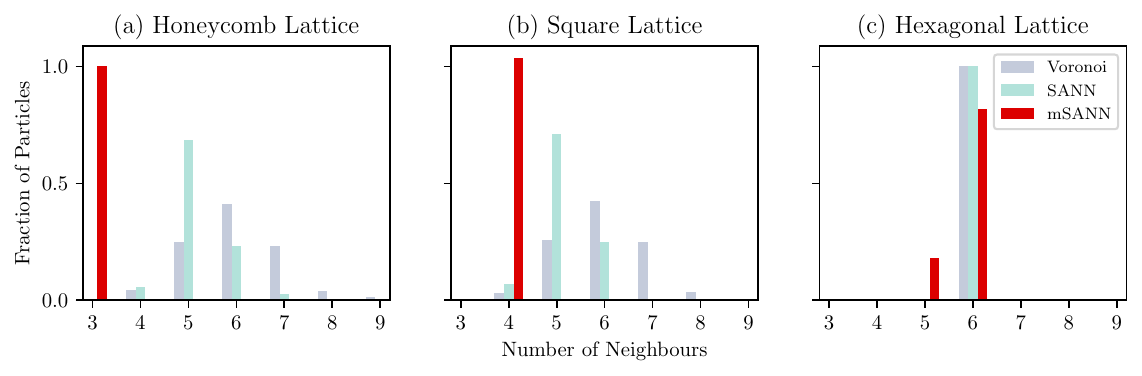}
\caption{Distributions of coordination numbers in 2D lattices identified using Voronoi, SANN, and mSANN: (a) honeycomb, (b) square, (c) hexagonal lattices corresponding to those shown in Fig.~\ref{fig:compare}. SANN and mSANN are computed without symmetrization.}
\label{fig:coord_lats}
\end{figure*}

\textbf{Two dimensions.}  
Representative neighbor networks are shown in Fig.~\ref{fig:compare}, and the resulting CN distributions in Fig.~\ref{fig:coord_lats}. For honeycomb and square lattices, Voronoi strongly overestimates the CN (peak at 6), while SANN reduces but does not eliminate this bias (peak at 5). By contrast, mSANN identifies the correct CN for all particles, yielding a sharp distribution. In the hexagonal lattice, Voronoi and SANN both recover CN=$6$ for all particles, whereas mSANN assigns CN=$5$ to about $20\%$ of the particles, arising from opposing thermal displacements of neighboring particles.

\begin{figure*}
\begin{ruledtabular}
\begin{tabular}{cccc}
 Lattice&Voronoi&SANN&mSANN \\ \hline
\ct{SC}&
\ct{\includegraphics[width=0.25\textwidth]{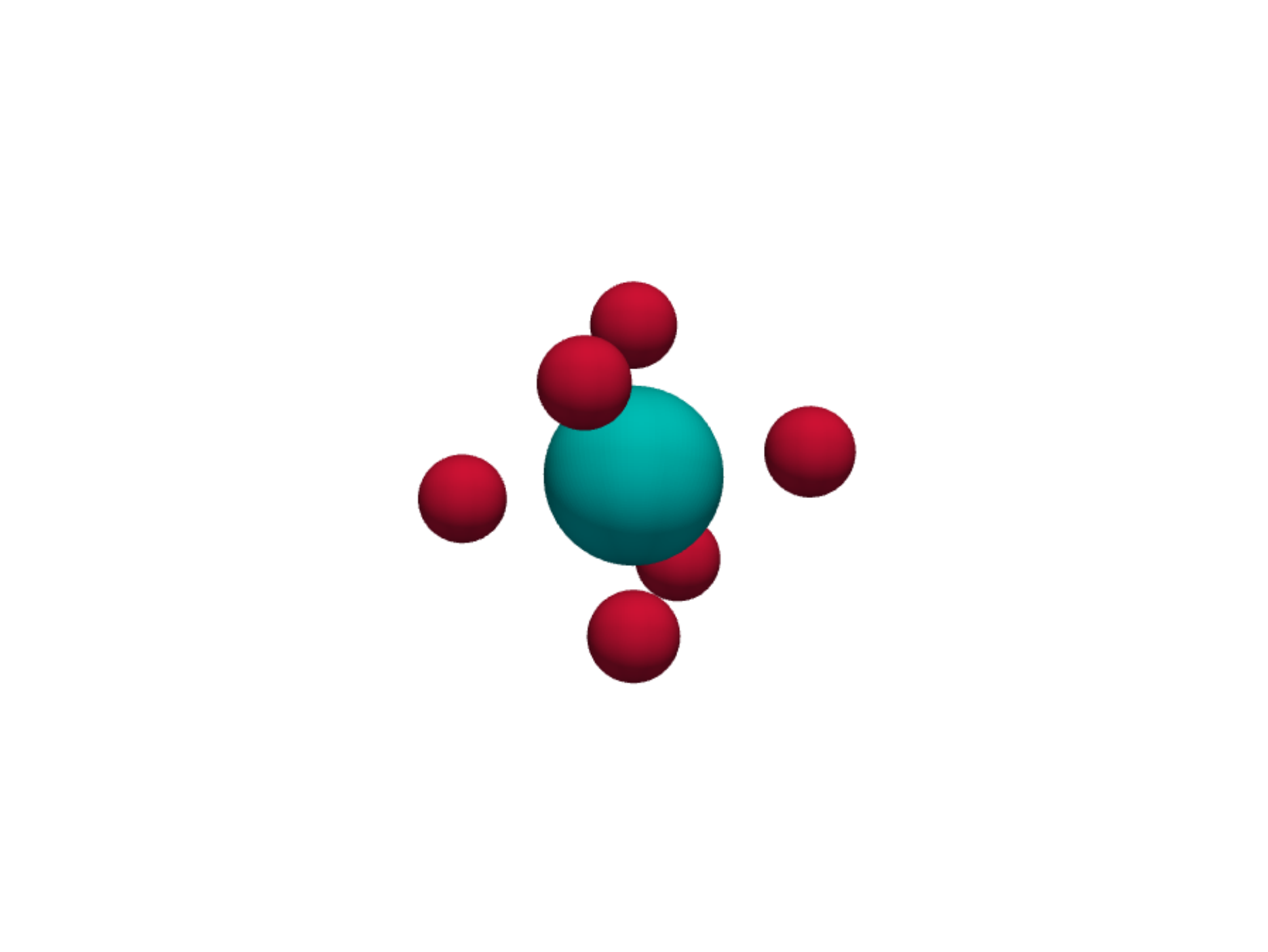}}&
\ct{\includegraphics[width=0.25\textwidth]{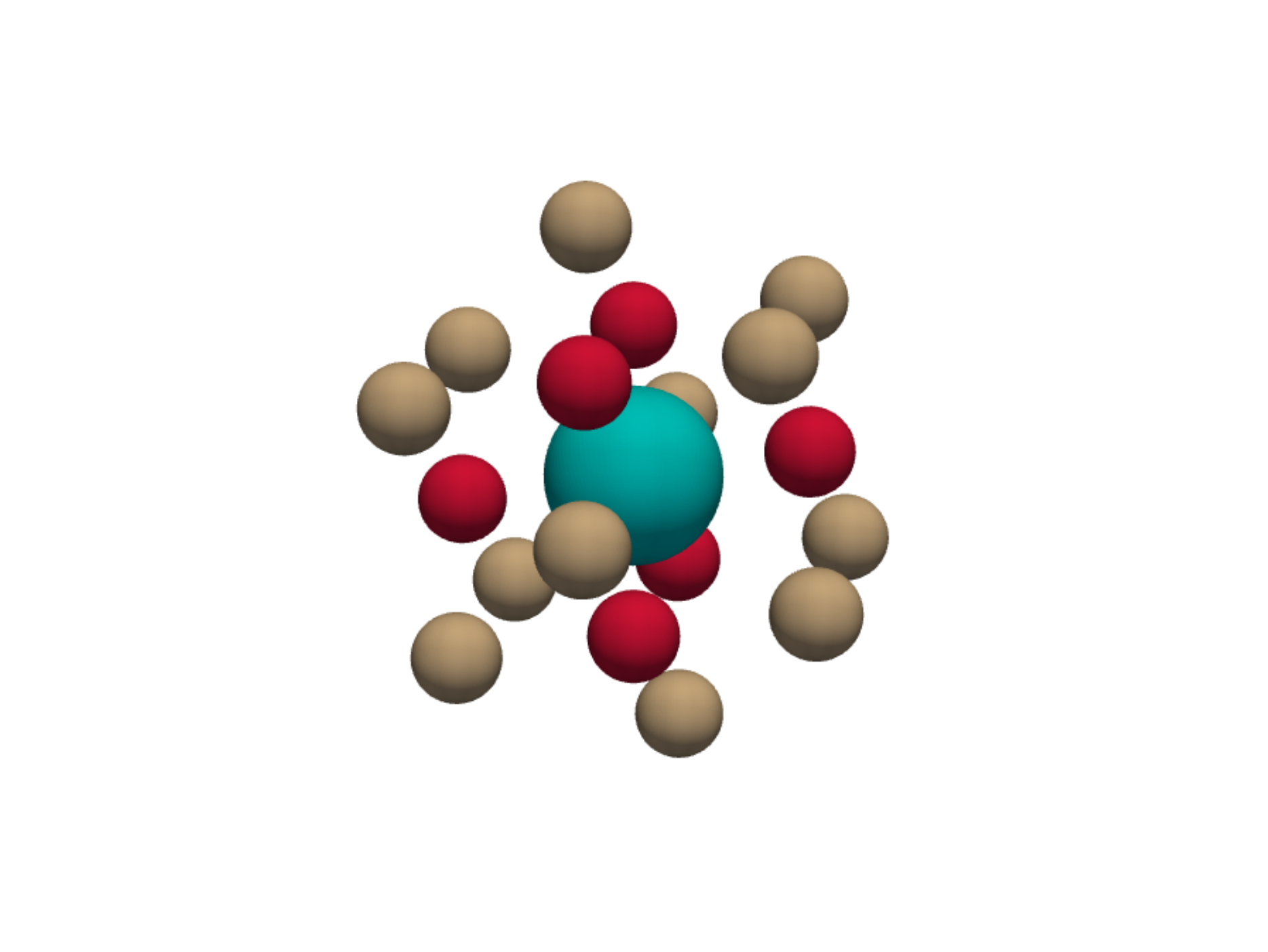}}&
\ct{\includegraphics[width=0.25\textwidth]{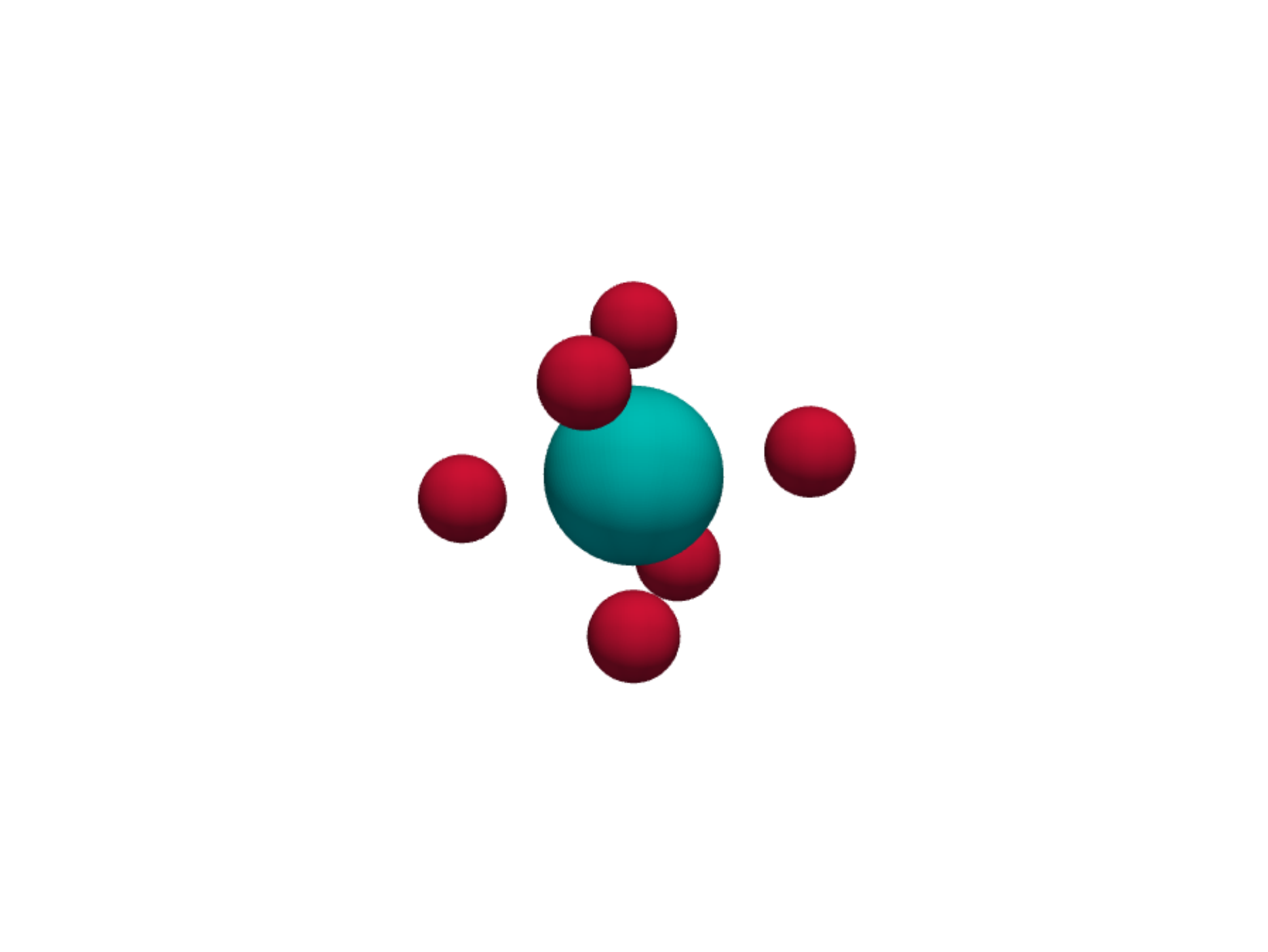}}
\\
\ct{BCC}&
\ct{\includegraphics[width=0.25\textwidth]{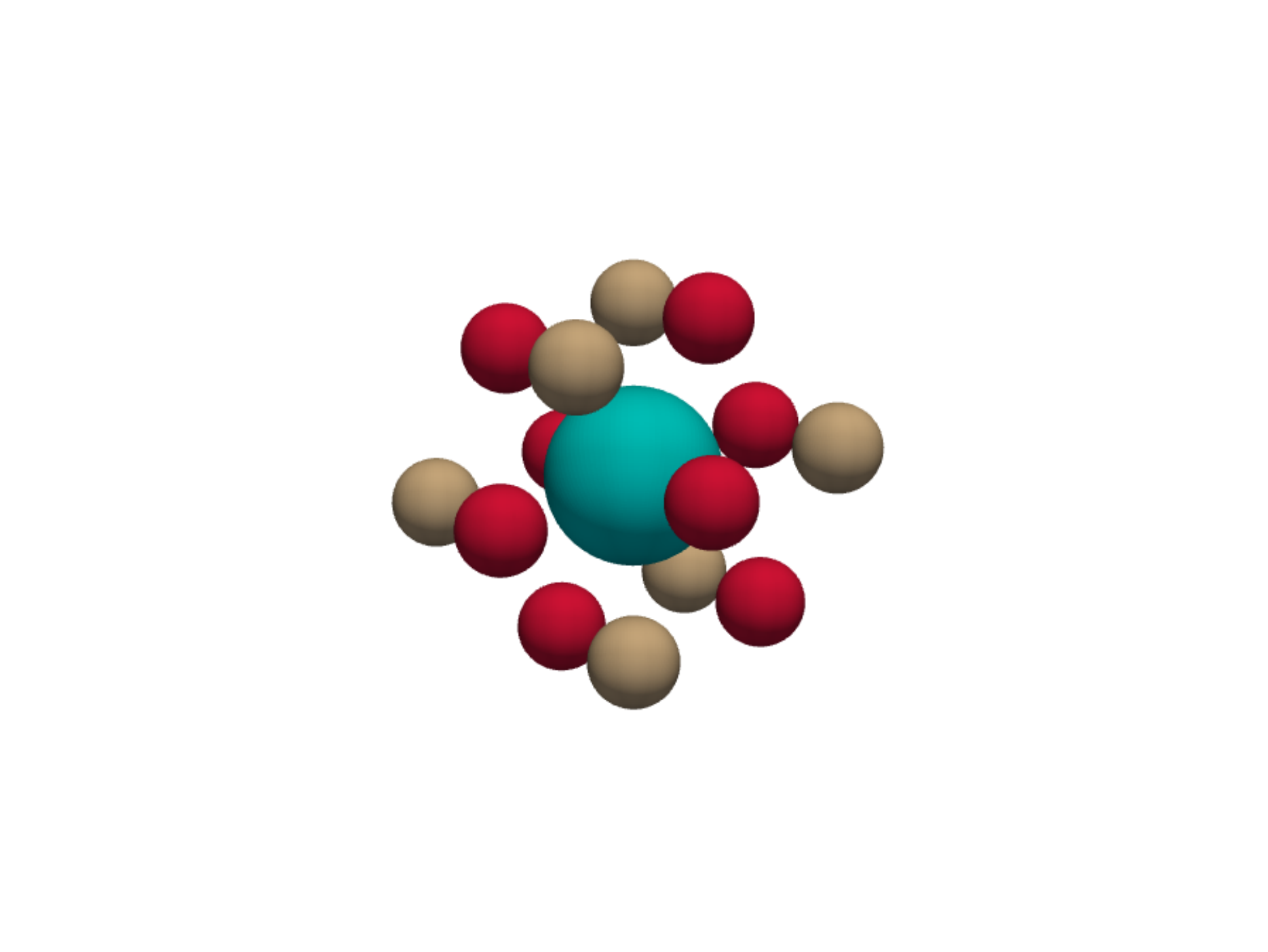}}&
\ct{\includegraphics[width=0.25\textwidth]{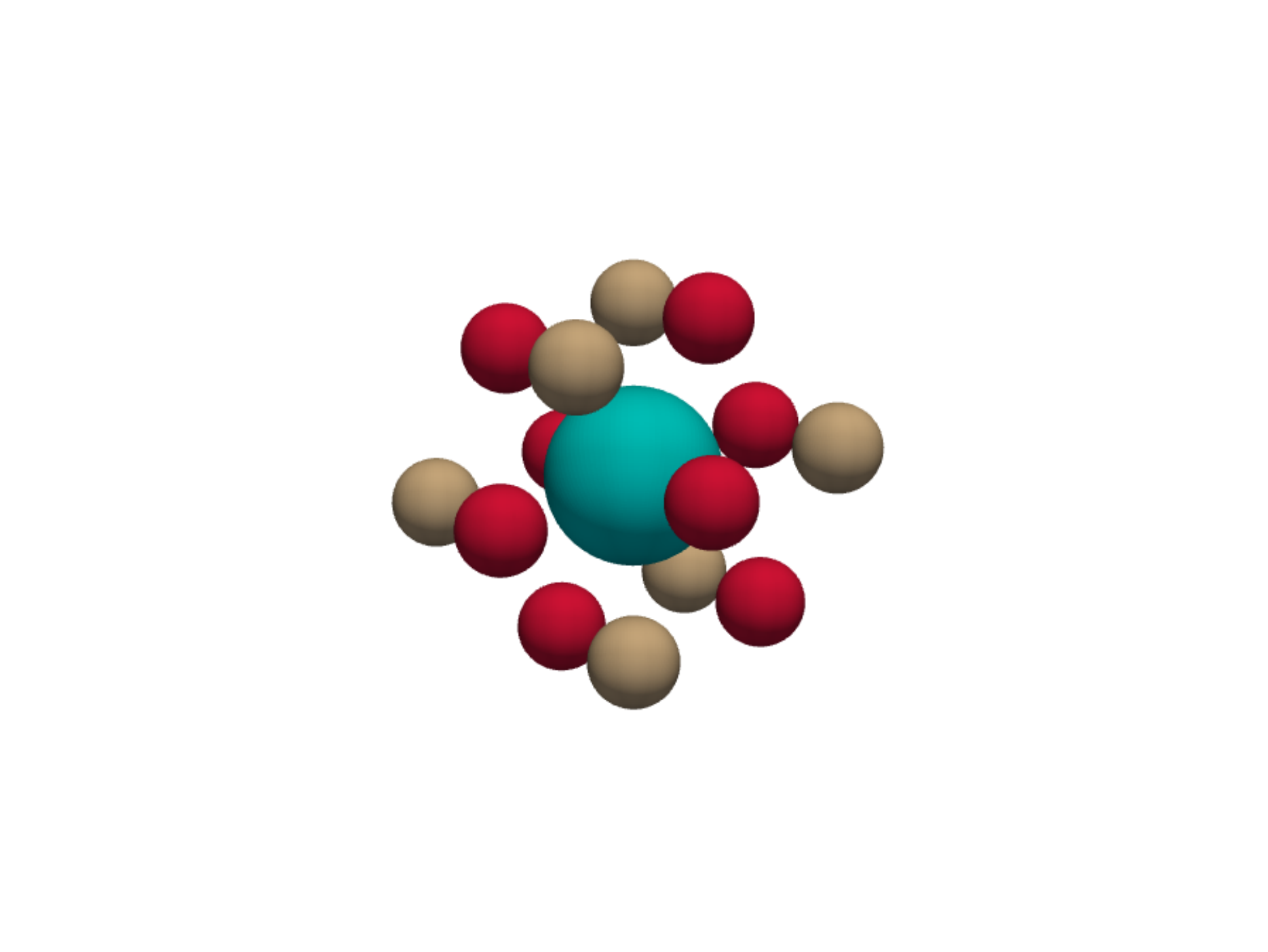}}&
\ct{\includegraphics[width=0.25\textwidth]{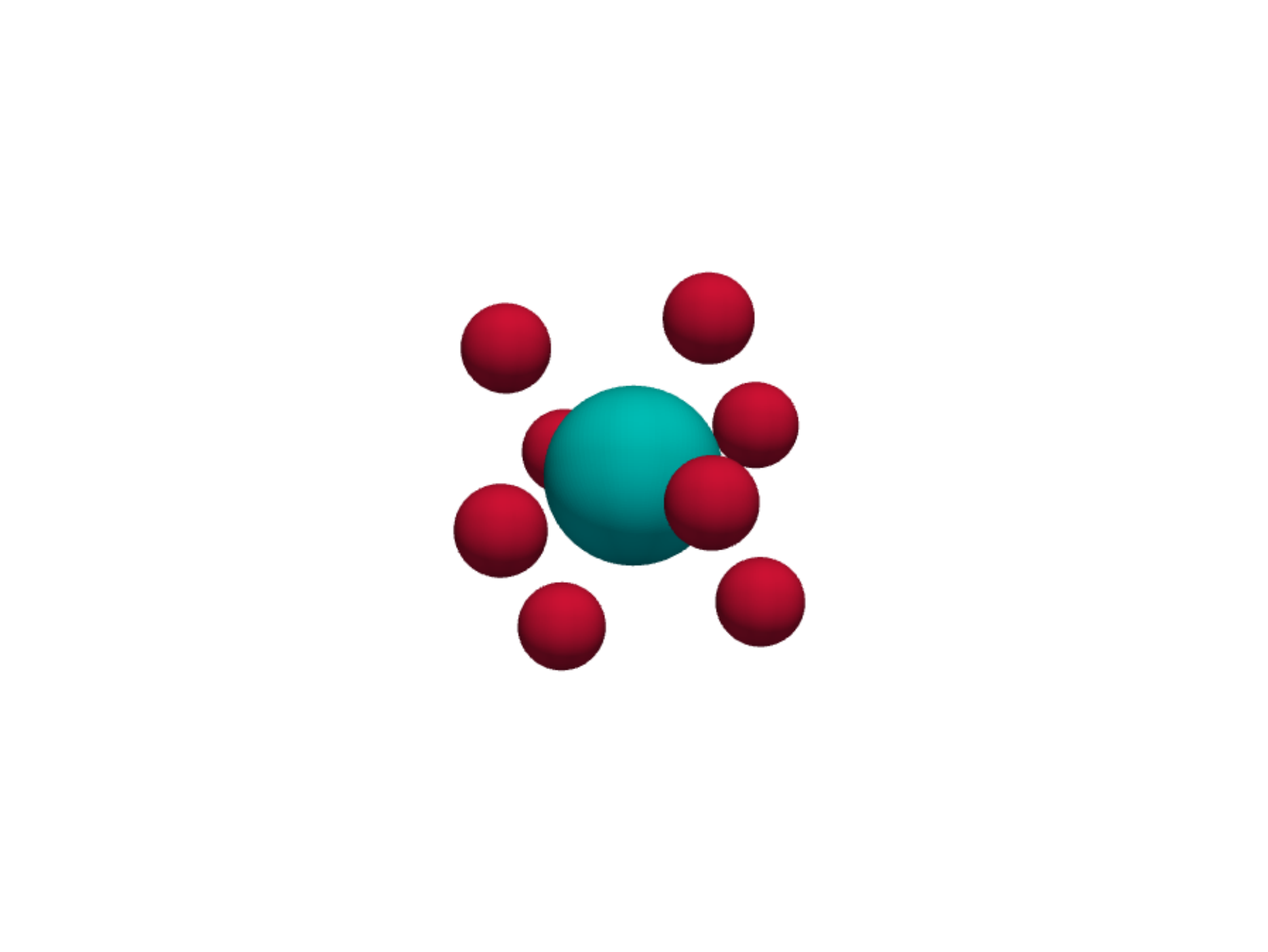}}
\\
\ct{FCC}&
\ct{\includegraphics[width=0.25\textwidth]{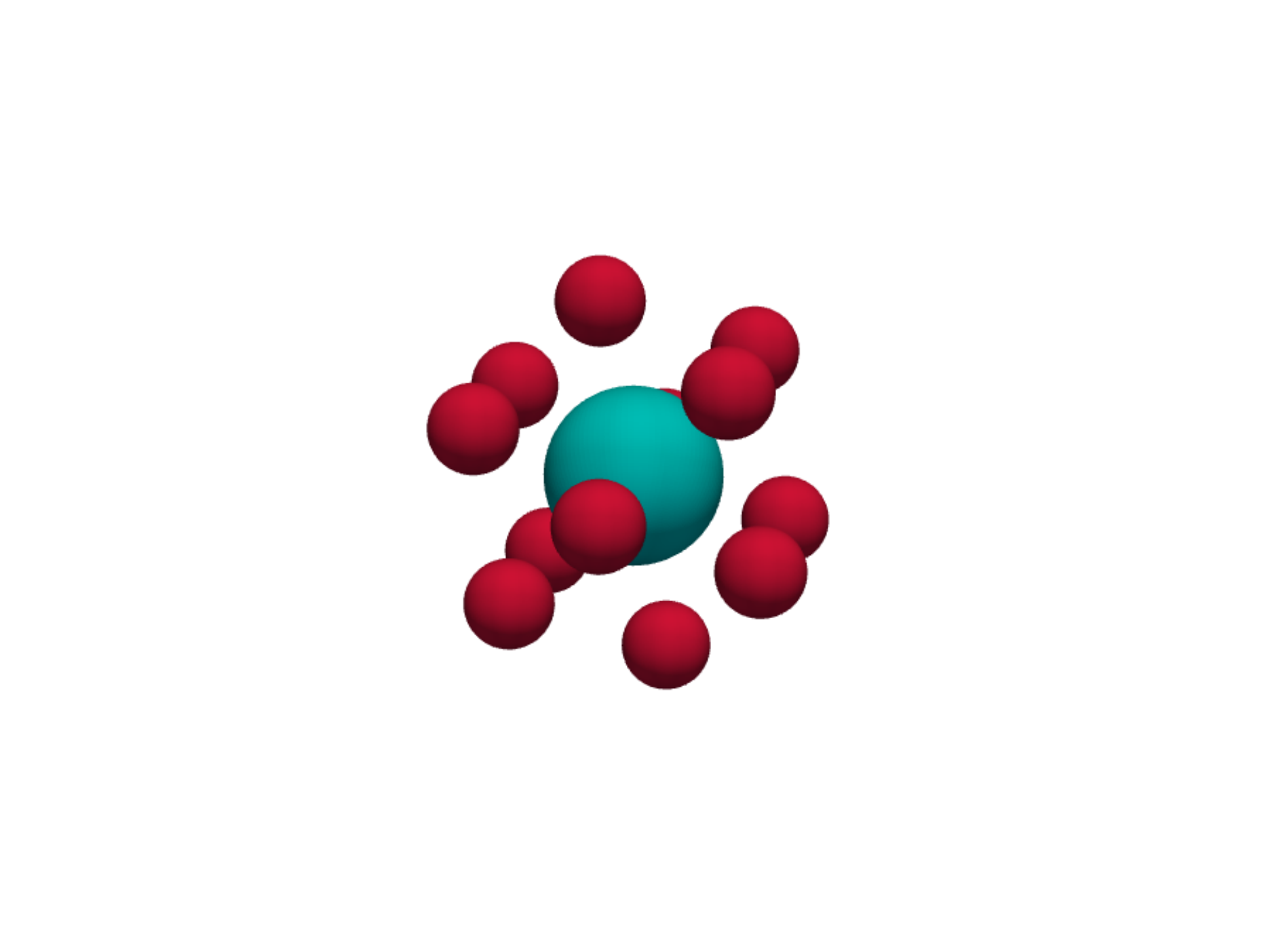}}&
\ct{\includegraphics[width=0.25\textwidth]{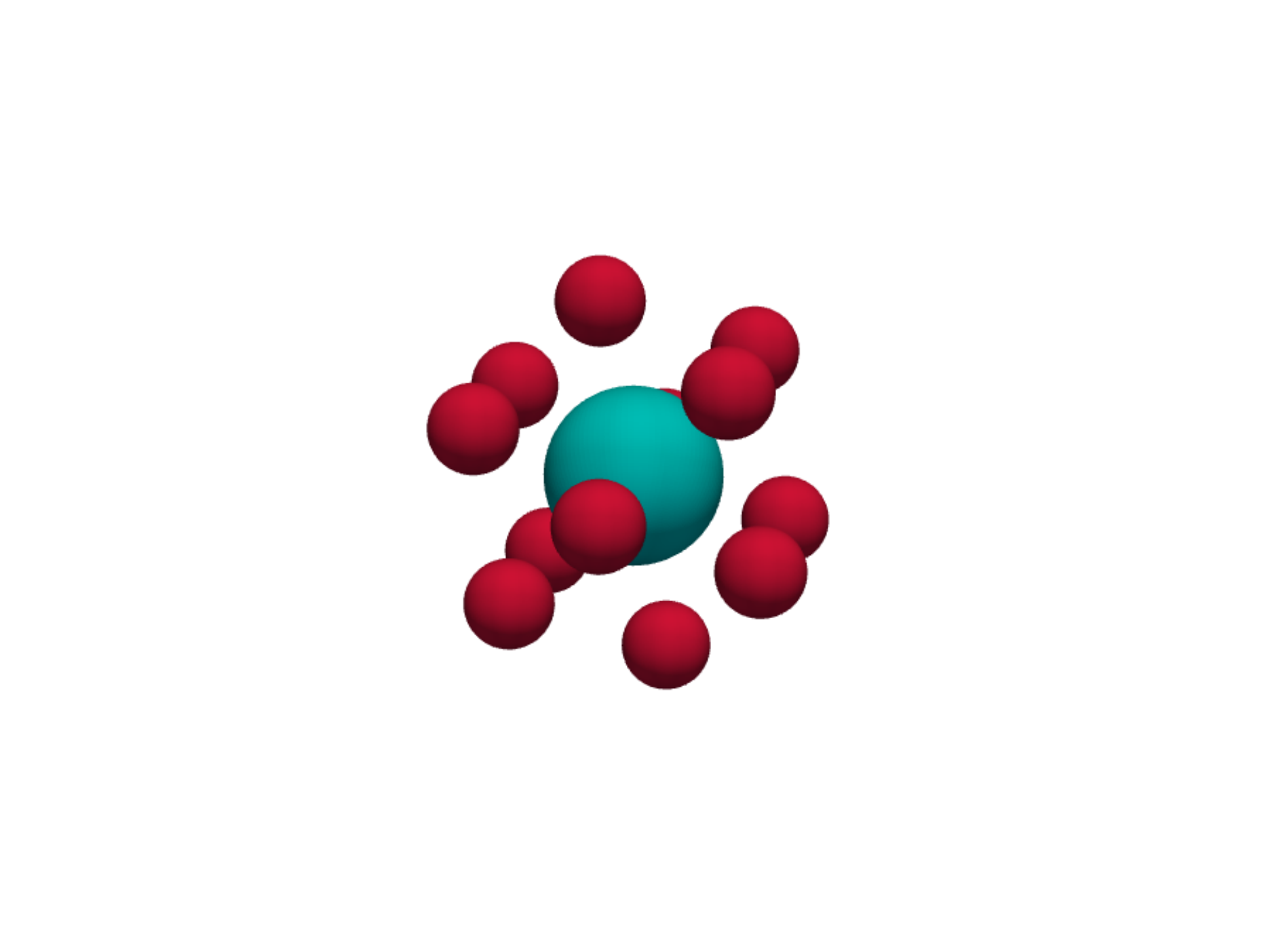}}&
\ct{\includegraphics[width=0.25\textwidth]{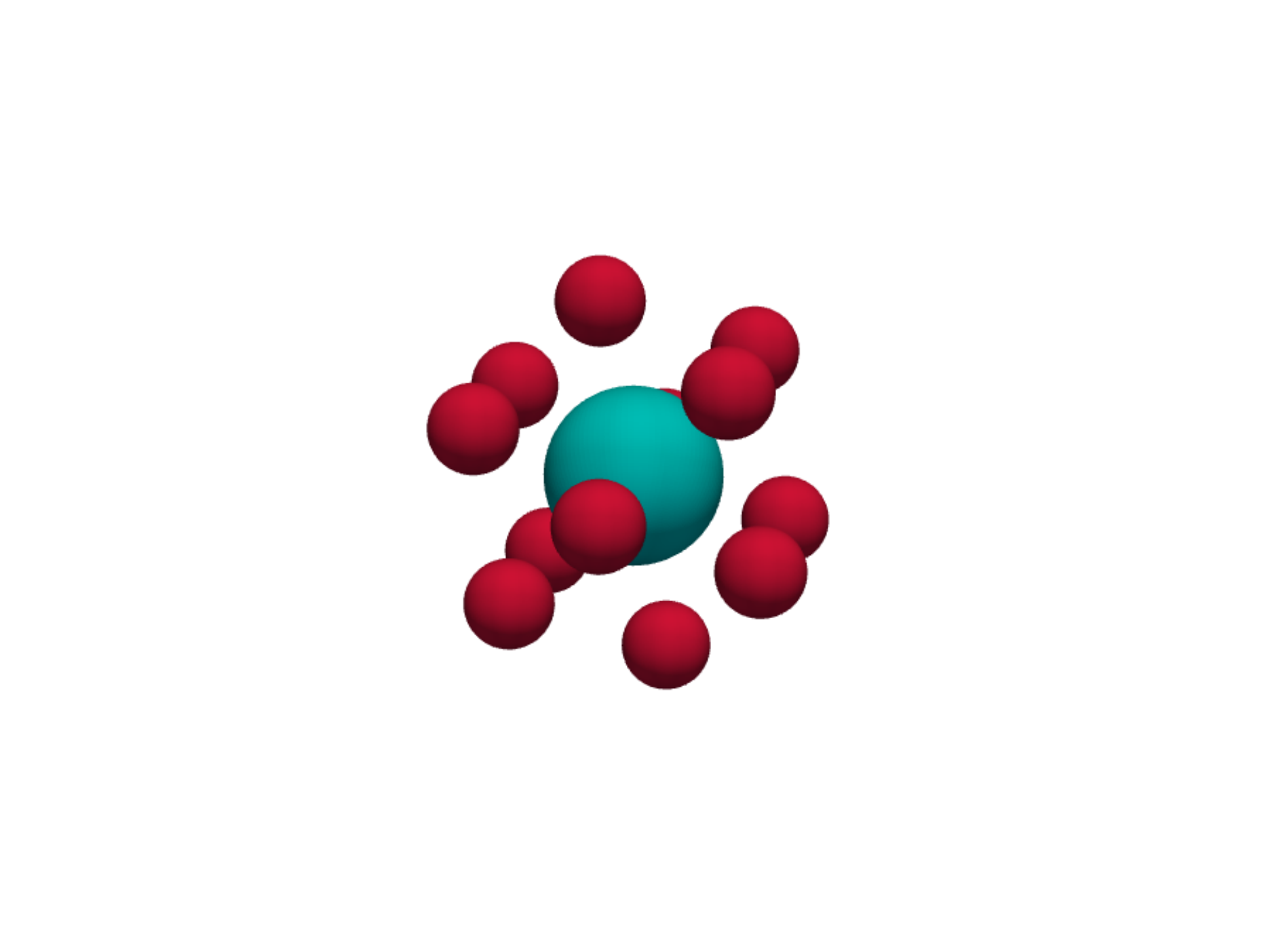}}
\\
\ct{HCP}&
\ct{\includegraphics[width=0.25\textwidth]{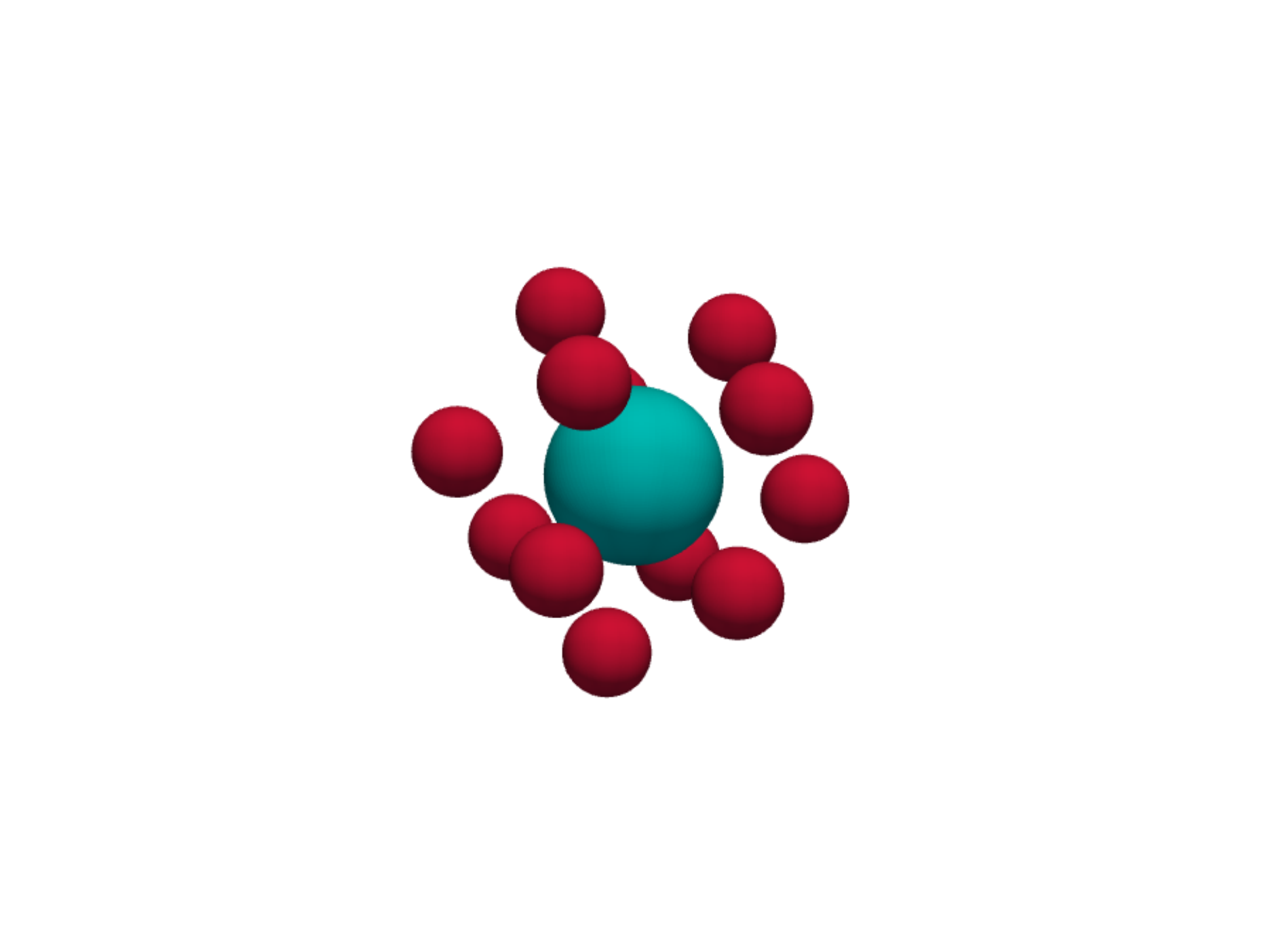}}&
\ct{\includegraphics[width=0.25\textwidth]{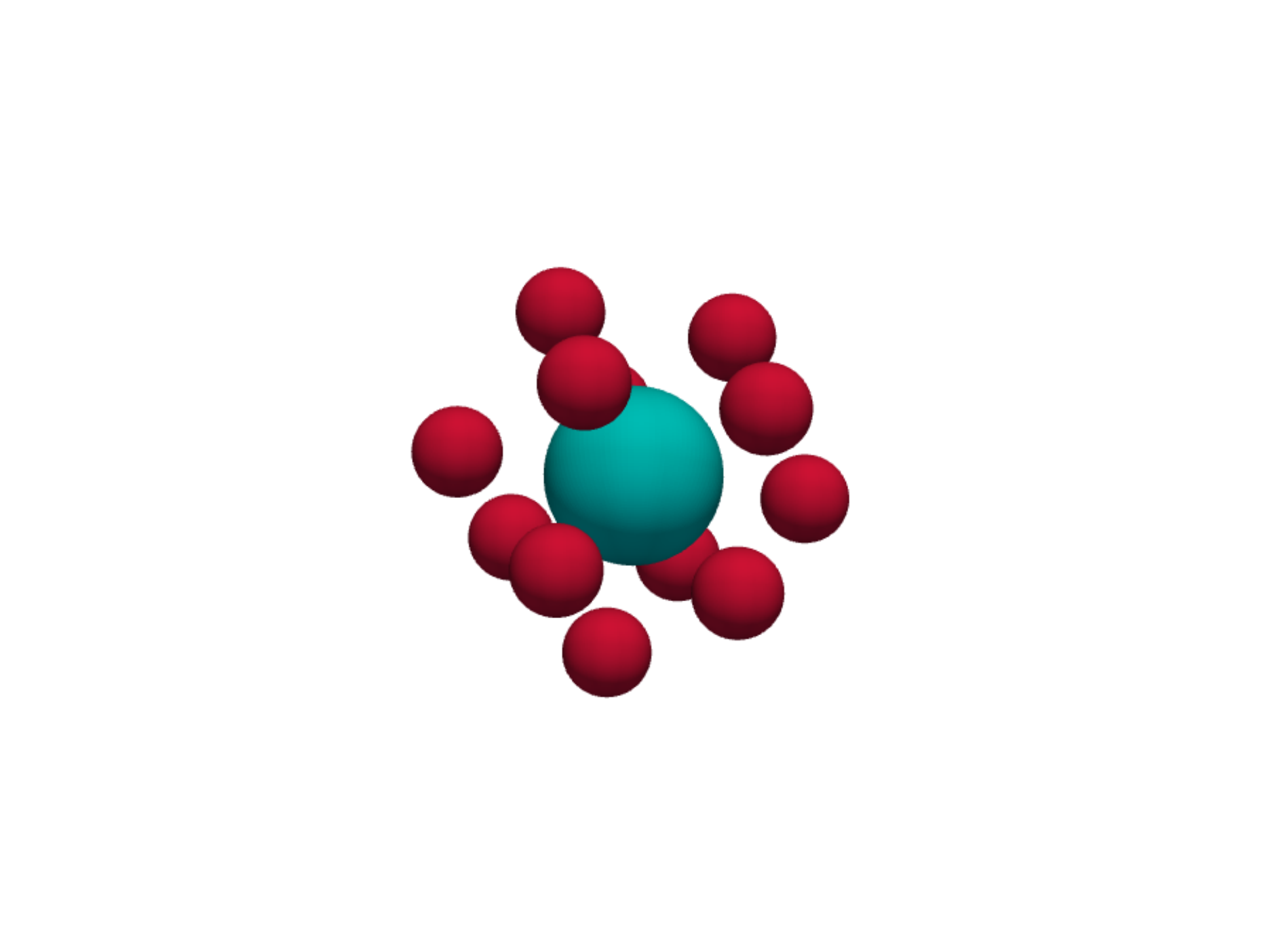}}&
\ct{\includegraphics[width=0.25\textwidth]{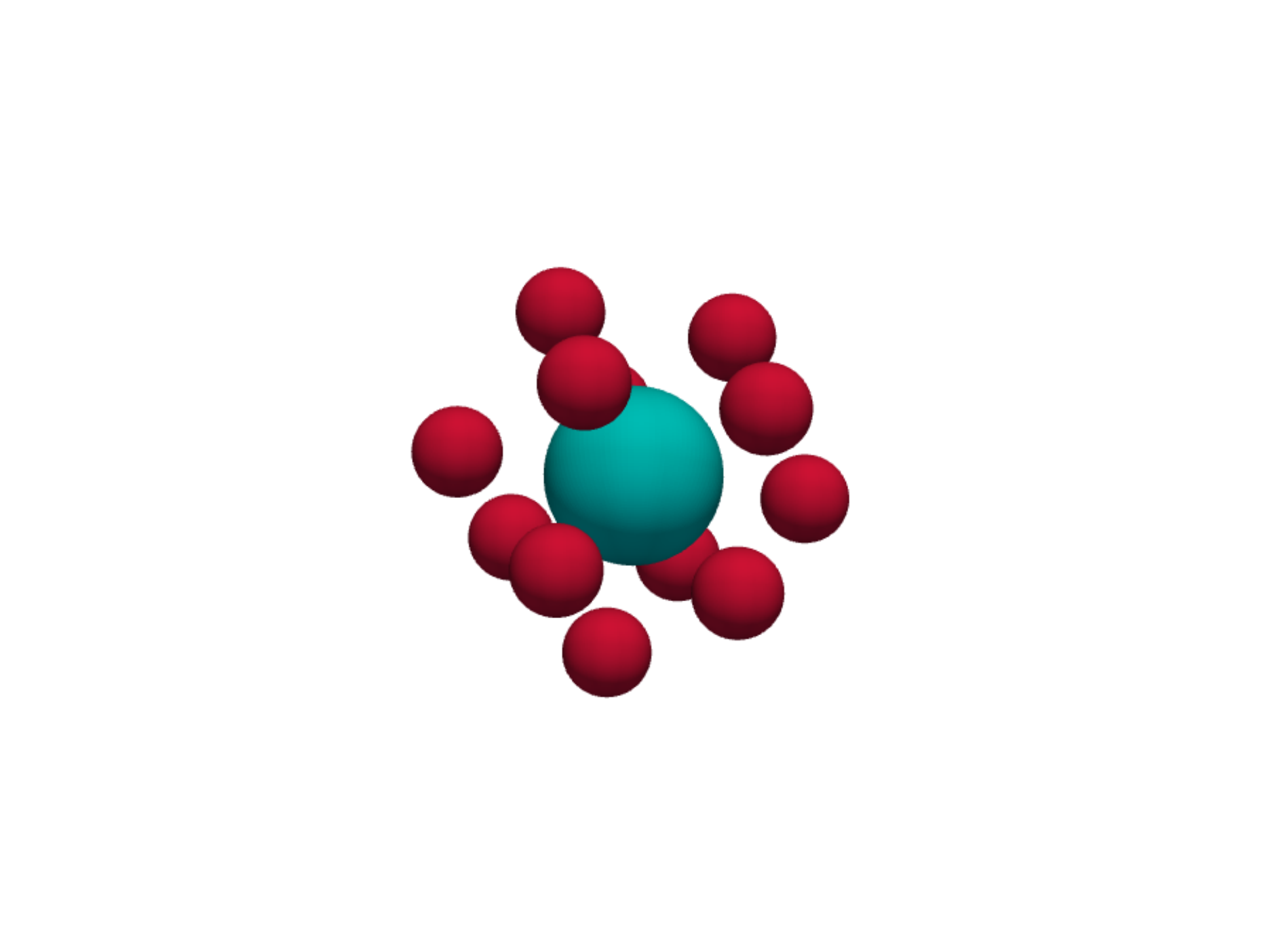}}
\\
\ct{Cubic Diamond}&
\ct{\includegraphics[width=0.25\textwidth]{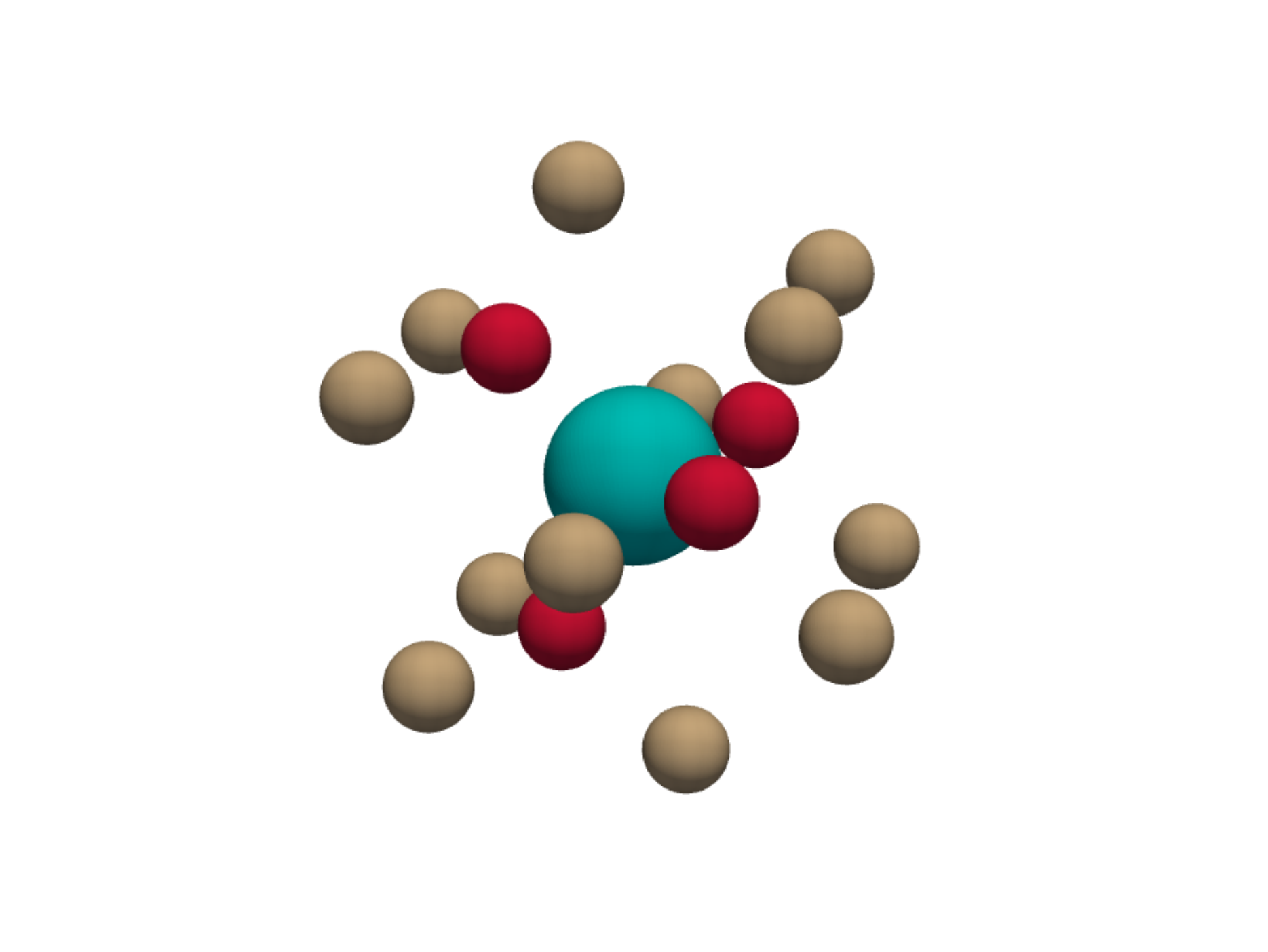}}&
\ct{\includegraphics[width=0.25\textwidth]{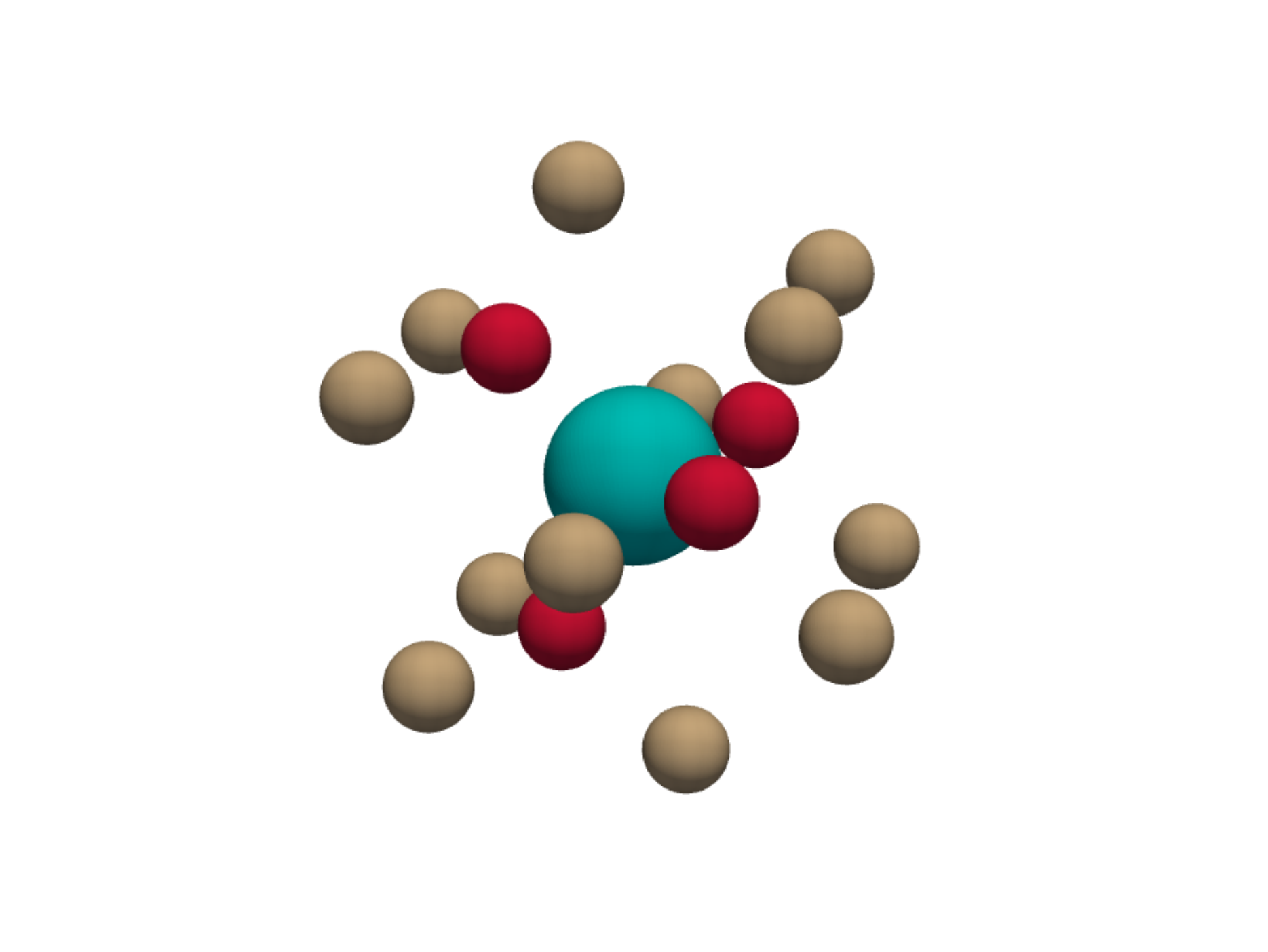}}&
\ct{\includegraphics[width=0.25\textwidth]{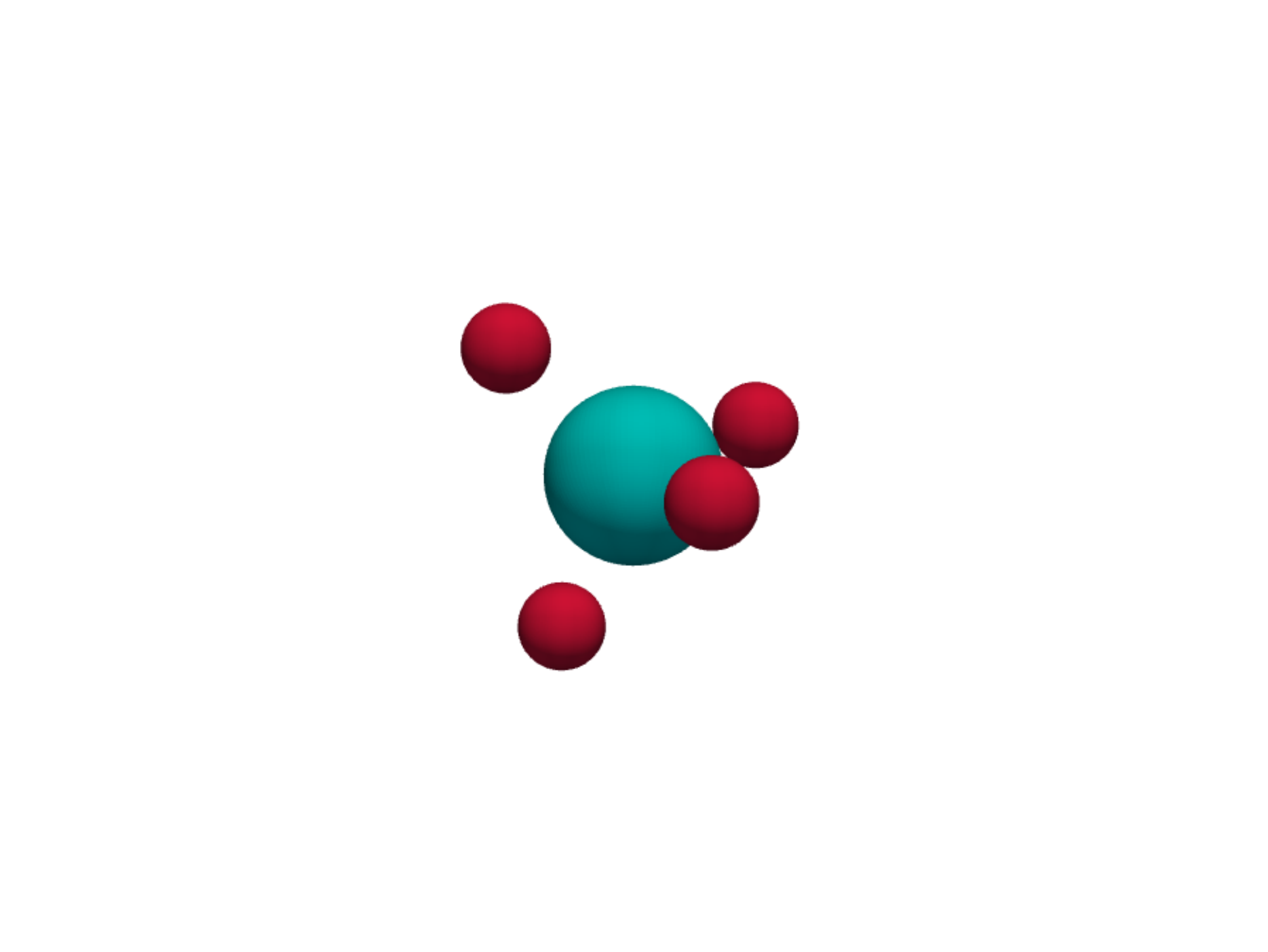}}
\\
\ct{$\alpha$-Graphite}&
\ct{\includegraphics[width=0.25\textwidth]{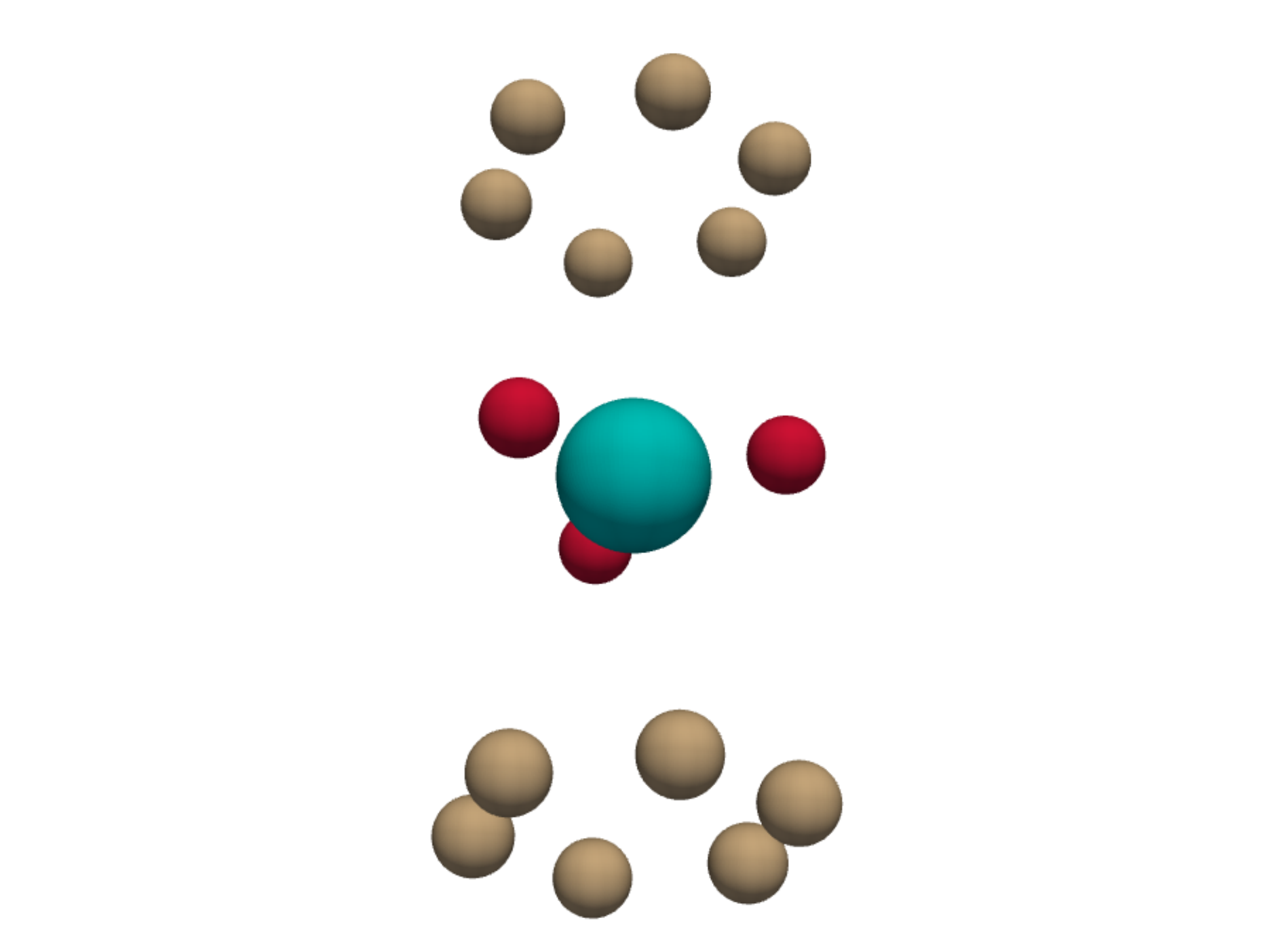}}&
\ct{\includegraphics[width=0.25\textwidth]{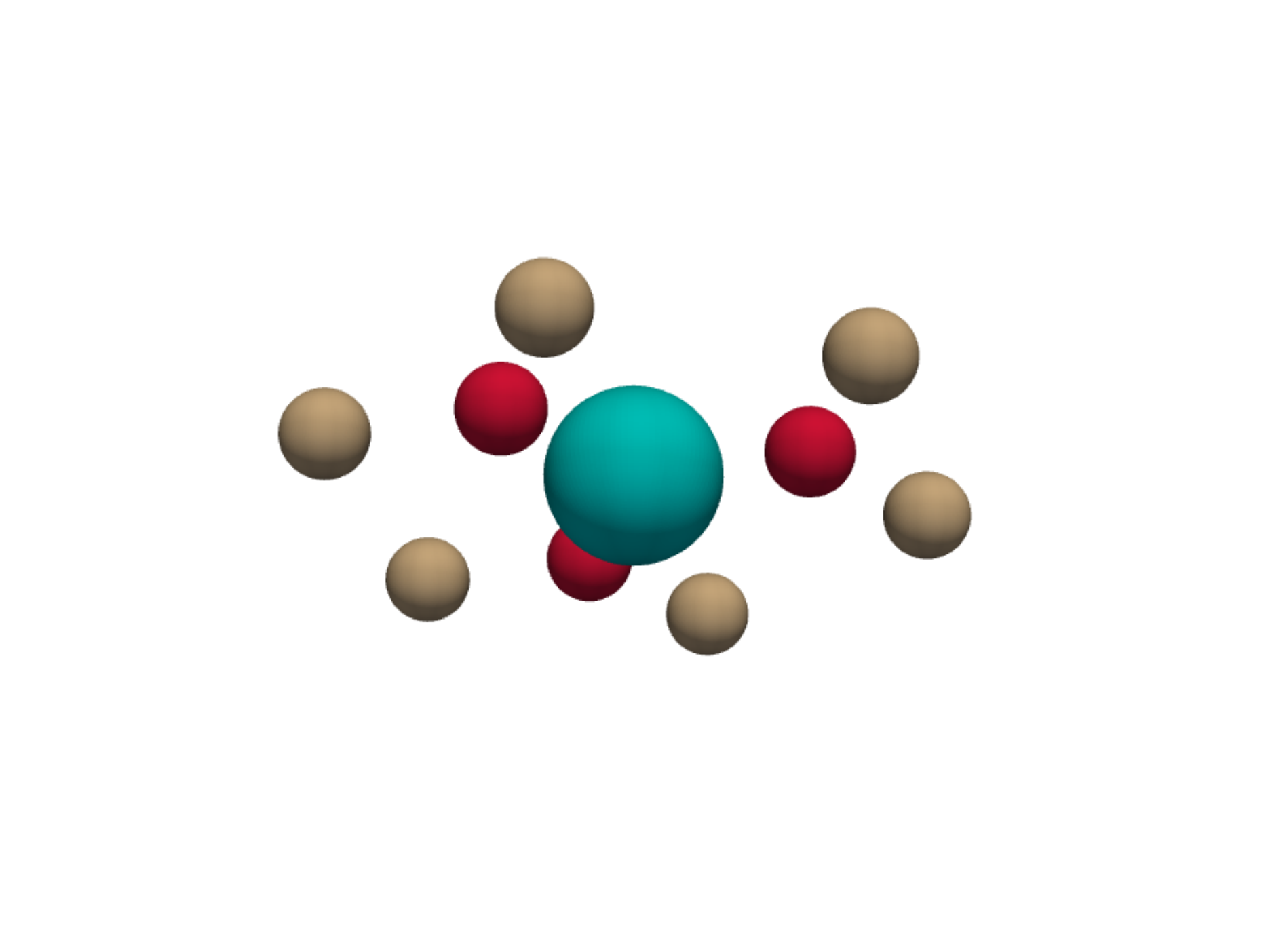}}&
\ct{\includegraphics[width=0.25\textwidth]{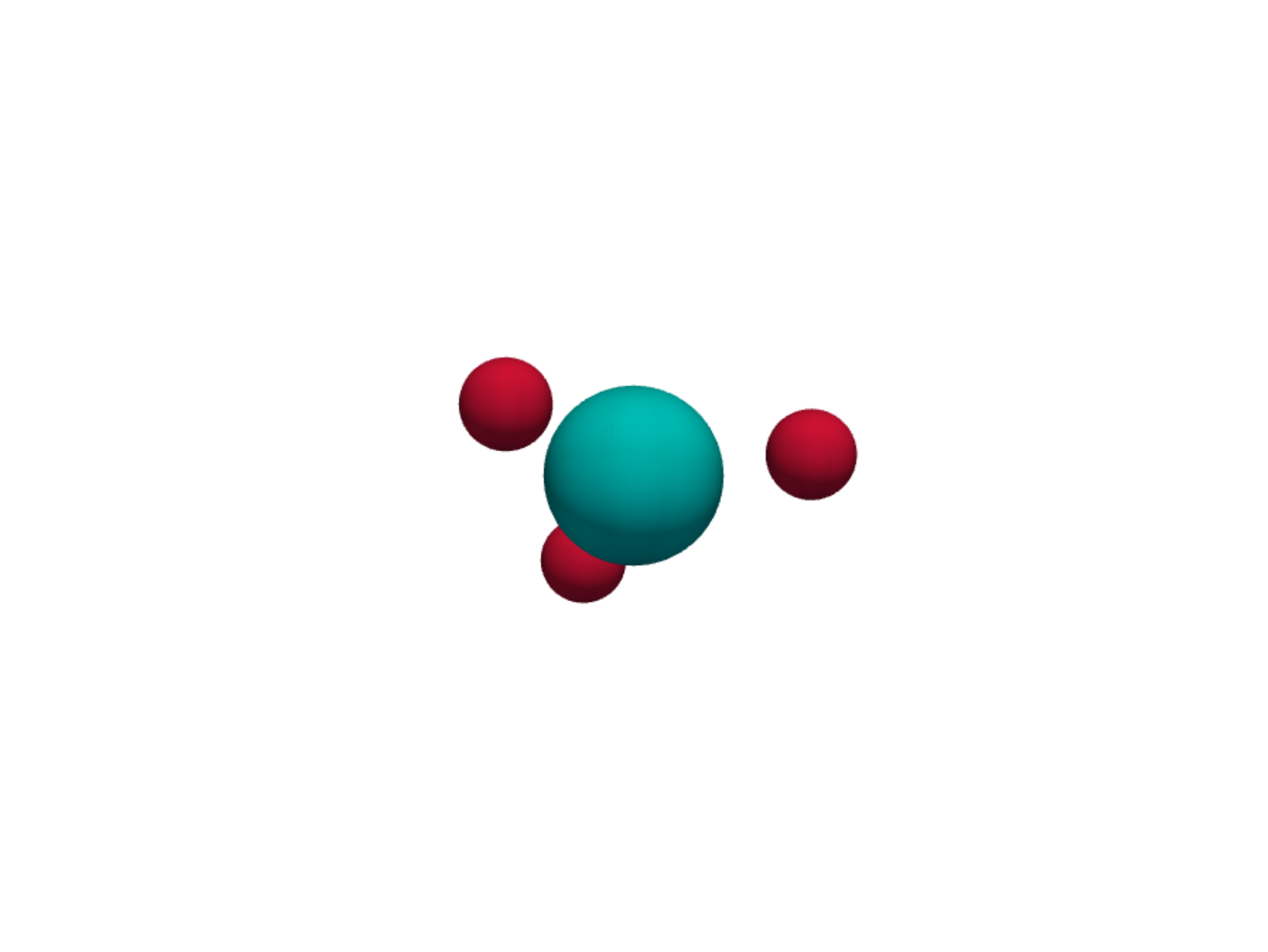}}
\\
\end{tabular}
\end{ruledtabular}
\caption{\label{fig:compare3d}Comparison of nearest-neighbor identification in ideal 3D lattices using Voronoi, SANN, and mSANN. Turquoise spheres indicate the central particle, crimson spheres its identified nearest neighbors (scaled by $0.5$), and tan spheres are the identified beyond nearest neighbors (scaled by $0.5$).}
\end{figure*}

\begin{figure*}
\includegraphics[width=\textwidth]{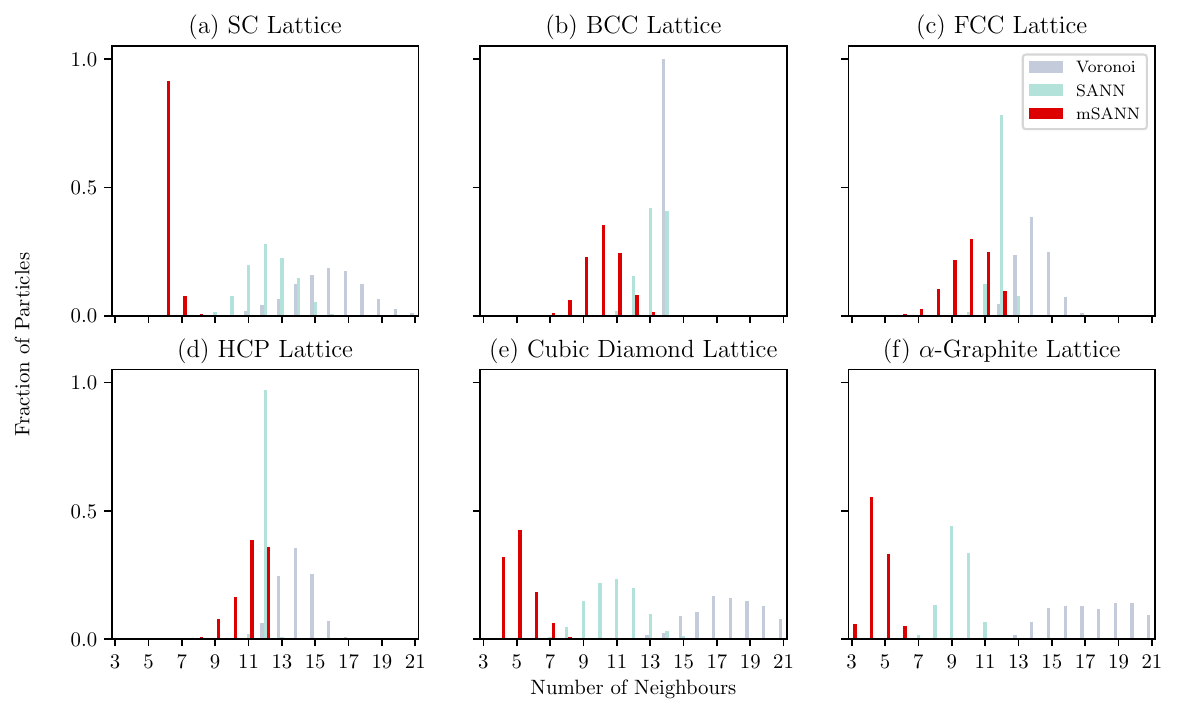}
\caption{Distributions of coordination numbers in 3D lattices identified using Voronoi, SANN, and mSANN. Systems correspond to those shown in Fig.~\ref{fig:compare3d}. SANN and mSANN are computed without symmetrization.}
\label{fig:coord_lats3d}
\end{figure*}

\textbf{Three dimensions.}  
For ideal lattices (Fig.~\ref{fig:compare3d}), Voronoi and SANN coincide except for SC and $\alpha$-graphite. In SC, Voronoi yields the correct CN while SANN overestimates. In $\alpha$-graphite, Voronoi correctly identifies in-plane neighbors but also includes spurious interplane neighbors, while SANN includes both nearest and next-nearest neighbors. In contrast, mSANN recovers the correct CN for all six structures.

Adding Gaussian noise broadens the distributions (Fig.~\ref{fig:coord_lats3d}). For low-CN lattices (SC, BCC, cubic diamond, $\alpha$-graphite), mSANN remains closest to the expected CN, confirming its correction of the overcounting problem. For high-CN lattices (FCC, HCP), SANN gives the most consistent results, indicating superior robustness under thermal fluctuations. Voronoi shows broad distributions in all but the BCC lattice, where it achieves its highest consistency.

\textbf{Summary.}  
Across bulk phases, mSANN consistently outperforms Voronoi and SANN for lattices with low coordination, while standard SANN remains preferable for highly coordinated structures subject to thermal fluctuations.

\subsection{12-fold symmetric quasicrystal approximants}

To further test the algorithms in two dimensions, we consider phases with multiple coordination numbers, namely 12-fold symmetric quasicrystals (QC12). These structures consist of squares and triangles and naturally exhibit CNs of $4$, $5$, and $6$, providing a benchmark for environments with fluctuating local density.

We analyze two realizations of the approximant: (i) a Stampfli tiling \cite{stampfli1986dodecagonal}  and (ii) a random square-triangle tiling of $1224$ particles. The Stampfli configuration was generated as an ideal tiling of $836$ vertices, which was then decorated with a binary mixture of additive hard disks of size ratio $0.4$, where the large disks sit at the vertices of the tiling, while the small disks sit at the centers of the squares, adding up to $1224$ particles in total. This configuration was then equilibrated at a packing fraction $0.824$. The random square–triangle tiling was obtained by applying zipper moves\cite{oxborrow1993random} to the Stampfli configuration, followed by the same equilibration protocol.

\begin{figure*}

\begin{ruledtabular}
\begin{tabular}{cccc}
 Tiling&Voronoi&SANN&mSANN
\\ \hline
\ct{Stampfli}&
\ct{\includegraphics[width=0.29\textwidth]{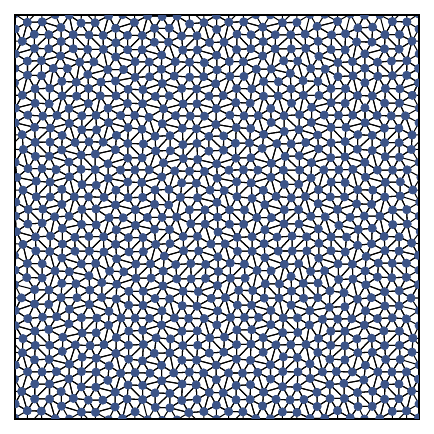}}&
\ct{\includegraphics[width=0.29\textwidth]{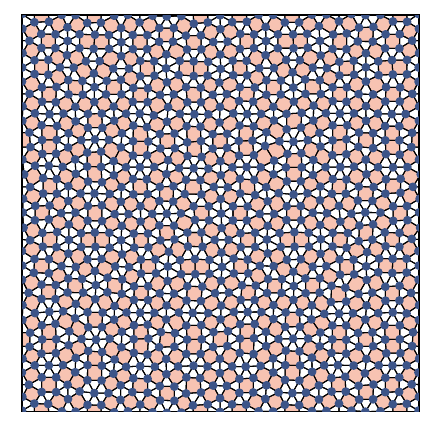} }&
\ct{\includegraphics[width=0.29\textwidth]{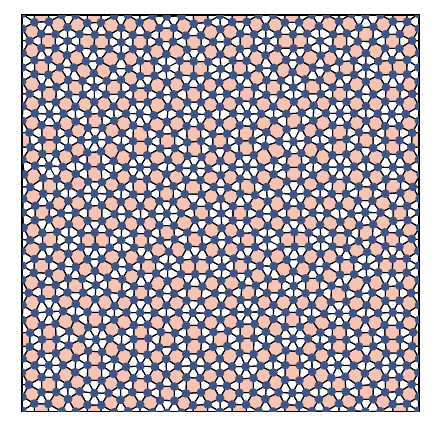}}
\\
\ct{Random}&
\ct{\includegraphics[width=0.29\textwidth]{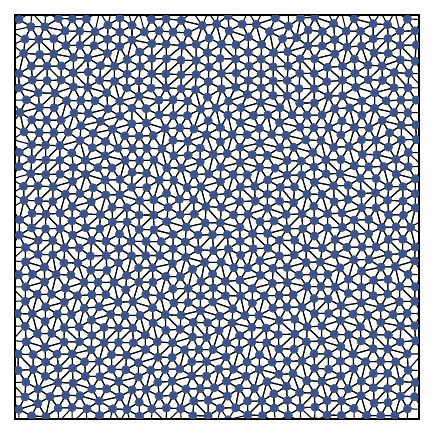}}&
\ct{\includegraphics[width=0.29\textwidth]{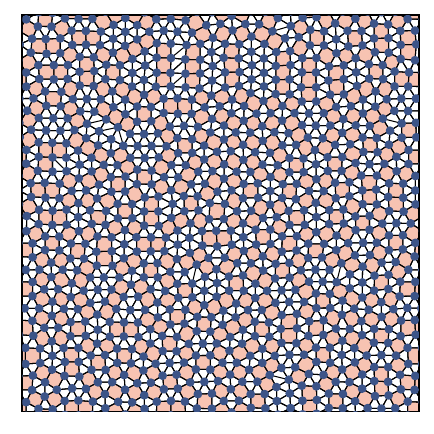} }&
\ct{\includegraphics[width=0.29\textwidth]{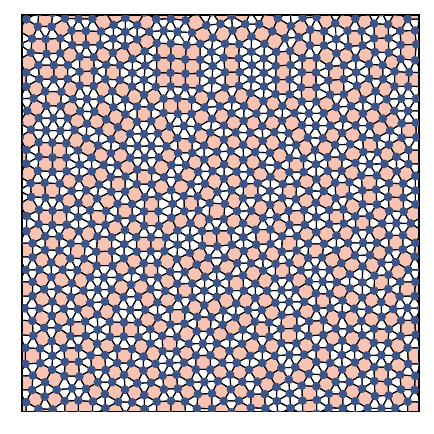}}
\\
\end{tabular}
\end{ruledtabular}
\caption{\label{fig:compareQC}Side-by-side comparison of nearest-neighbor identification using Voronoi construction, SANN  and mSANN algorithms on 12-fold symmetric quasicrystal approximants. Dots represent particles and line segments connect identified neighbors. All lattices are simulated, thus, they include correlated thermal noise. Additive symmetrization is used for the visualization of SANN and mSANN.}
\end{figure*}

\begin{figure*}
\includegraphics[width=\textwidth]{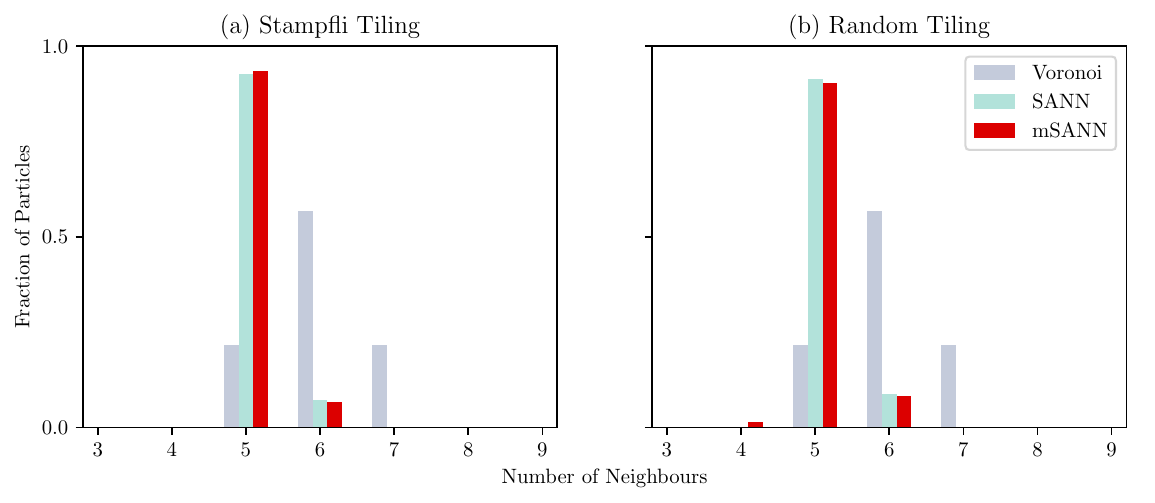}
\caption{The distributions of the number of nearest neighbors in a 12-fold symmetric (a) Stampfli-infilation , (b) random square-triangle tiling (depicted in Fig. \ref{fig:compareQC}) where nearest neighbors were identified by Voronoi construction, SANN, and mSANN. SANN and mSANN are computed without symmetrization.}
\label{fig:coord_qc}
\end{figure*}

All three algorithms were applied to both realizations. Representative results are shown in Fig.~\ref{fig:compareQC}, where identified square tiles are highlighted in orange for clarity, and the resulting CN distributions are given in Fig.~\ref{fig:coord_qc}.

First, we observe that Voronoi construction fails to identify any square tiles, which is a consequence of its intrinsic triangulation. By design, it cannot generate non-triangular polygons without additional post-processing. This is reflected in Fig.~\ref{fig:coord_qc}, where Voronoi consistently yields an average CN of $6$ for both approximants.

In contrast, both SANN and mSANN successfully recover square tiles. For the Stampfli tiling, the two algorithms give nearly identical results, including the distribution of CNs. Differences emerge in the randomized tilings, particularly in square-dominated regions: SANN often introduces spurious links across square diagonals, inflating the CN, whereas mSANN preserves the intended square topology.

These results highlight a key advantage of mSANN in complex environments. Unlike Voronoi, which is restricted to triangulations, and unlike SANN, which is prone to overcounting in low-CN motifs, mSANN consistently reproduces the intended local topology of quasicrystal approximants. This robustness across heterogeneous local environments suggests that mSANN is well suited for the analysis of disordered or aperiodic phases, where multiple coordination numbers coexist.

\subsection{Coexisting phases and defects}

As a final test case, we analyze two structures that include a phase coexistence: QC12-hexagonal and QC12-square. These systems combine multiple coordination numbers ($4$, $5$, and $6$) with interfaces and defects, providing a comprehensive benchmark for algorithm robustness.

Both coexistence states were generated by simulating square-shoulder particles in the canonical ensemble, with pair potential
\begin{equation}\label{eq:sq_shoulder}
V(r) = 
    \begin{cases} 
      \infty & r\leq \sigma, \\
      \epsilon & \sigma < r \leq \delta, \\
      0 & \delta < r,
   \end{cases}
\end{equation}
where $r$ is the interparticle distance, $\sigma$ the particle diameter, $\epsilon$ the shoulder height, and $\delta = 1.4\sigma$ the shoulder width. The phase behavior of this system was studied extensively and it is known to form 12-fold QC, hexagonal and square lattices \cite{dotera2014mosaic,pattabhiraman2015stability,pattabhiraman2017phase}. 

To generate the QC12-hexagonal coexistence, we start from a hexagonal lattice and replace its central region with a patch of the Stampfli tiling. Each vertex is then decorated with a square-shoulder particle with interaction strength $\epsilon/k_BT = 3.333$, resulting in a total of $N = 1865$ particles at a number density of $\rho\sigma^2 = 1.032$. The configuration was subsequently equilibrated using event-driven molecular dynamics (EDMD)\cite{smallenburg2022efficient}.

The QC12-square coexistence was prepared analogously, starting from a square lattice. The resulting system contained $N = 1481$ particles at a number density of $\rho\sigma^2 = 0.9542$, and was likewise equilibrated using EDMD\cite{smallenburg2022efficient}.

\begin{figure*}
\begin{ruledtabular}
\begin{tabular}{cccc}
 Coexistance&Voronoi&SANN&mSANN
\\ \hline
\ct{QC12-Hex}&
\ct{\includegraphics[width=0.29\textwidth]{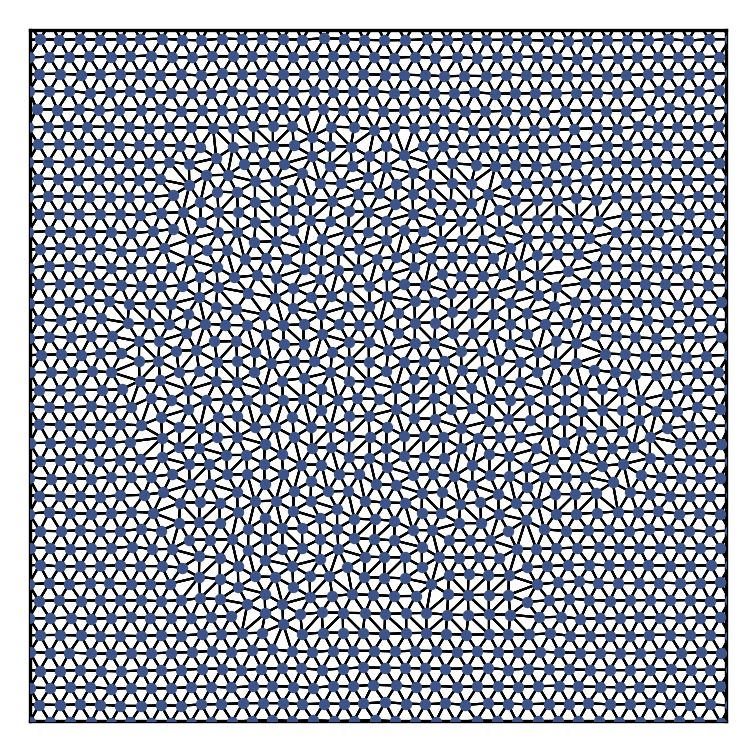}}&
\ct{\includegraphics[width=0.29\textwidth]{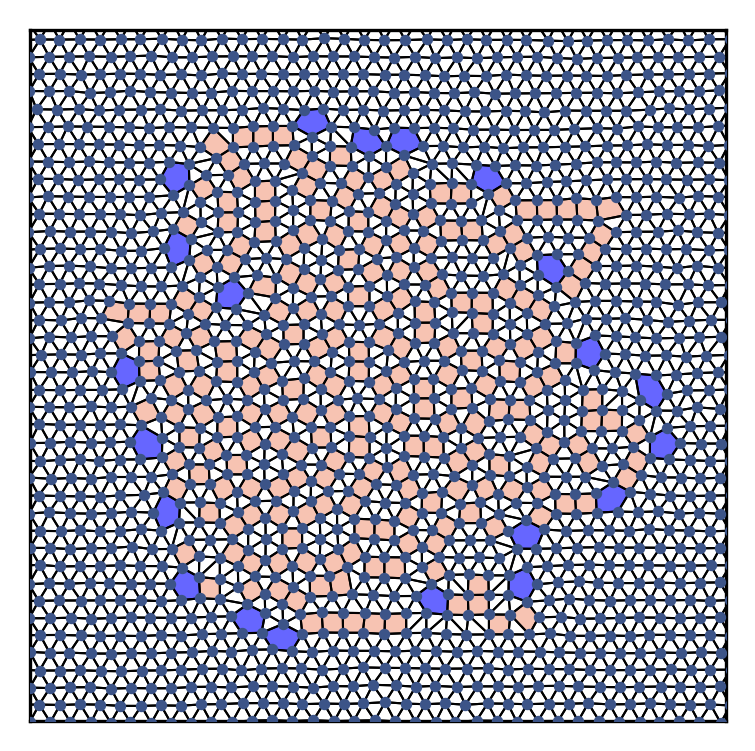} }&
\ct{\includegraphics[width=0.29\textwidth]{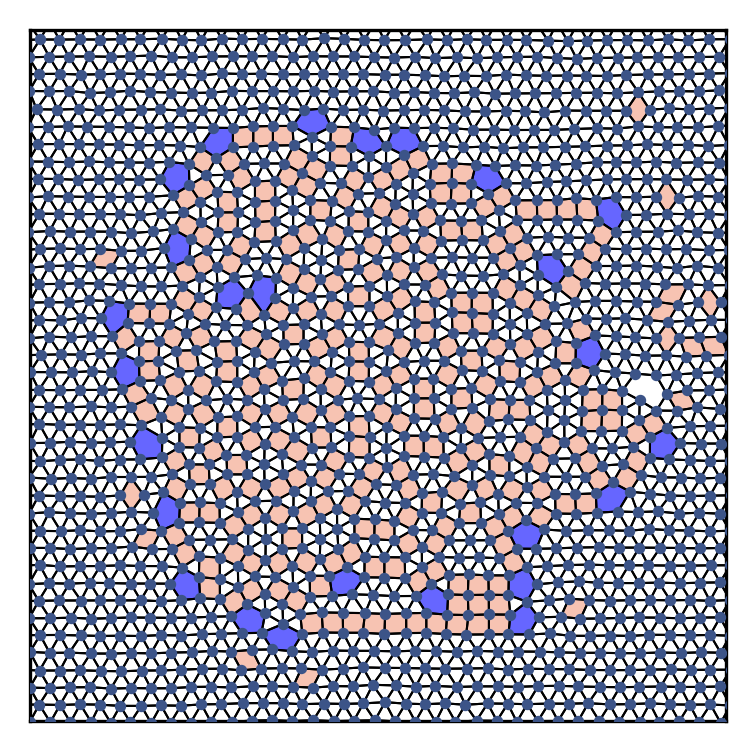}}
\\
\ct{QC12-Sq}&
\ct{\includegraphics[width=0.29\textwidth]{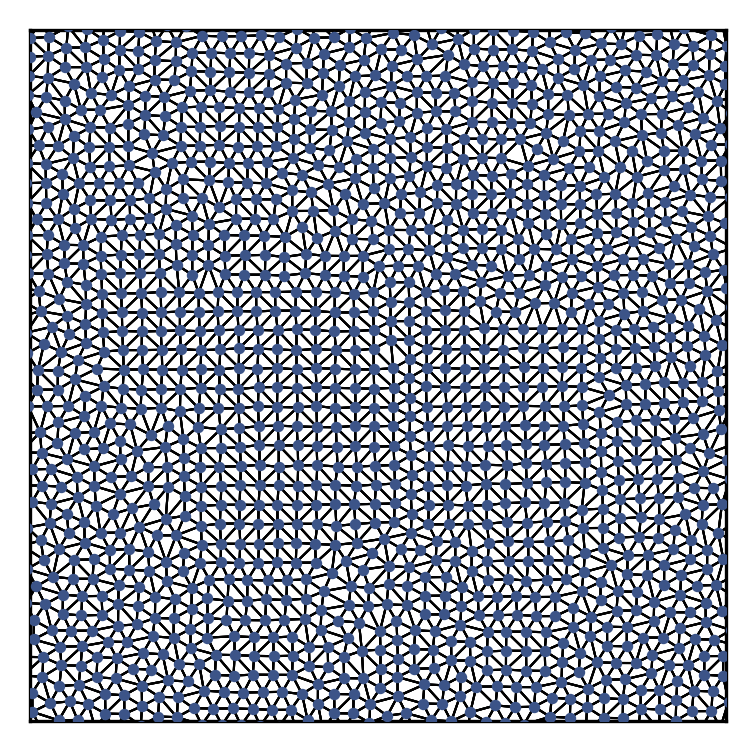}}&
\ct{\includegraphics[width=0.29\textwidth]{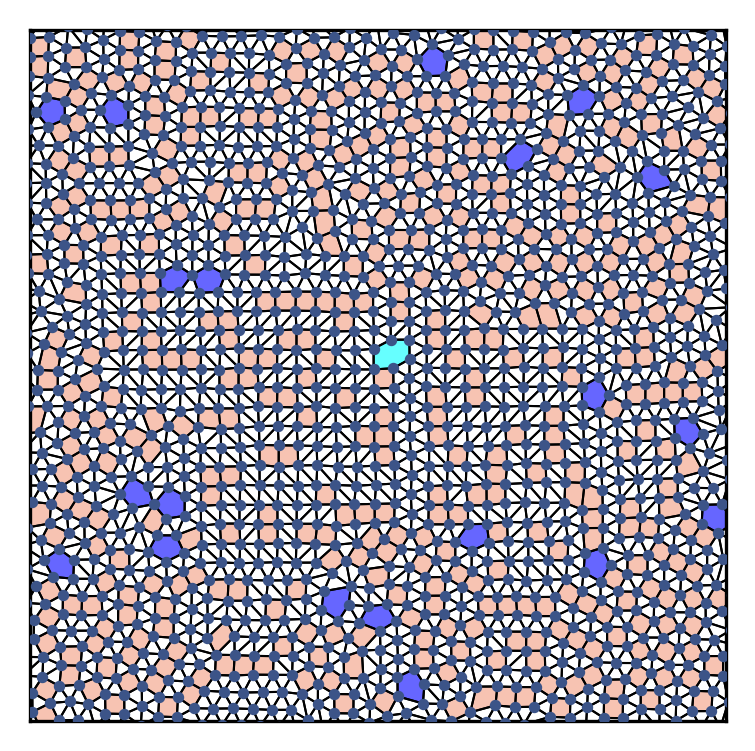} }&
\ct{\includegraphics[width=0.29\textwidth]{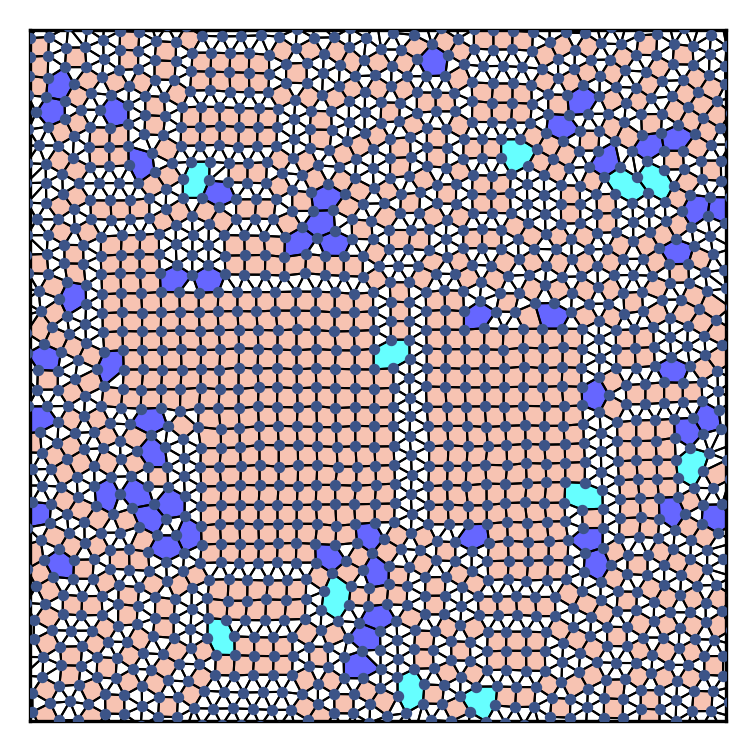}}
\\
\end{tabular}
\end{ruledtabular}
\caption{Side-by-side comparison of nearest-neighbor identification using Voronoi construction, SANN and mSANN algorithms on 12-fold symmetric quasicrystal coexisting with the hexagonal and the square phases. Additive symmetrization is used for the visualization of SANN and mSANN.}
\label{fig:coex}
\end{figure*}

As seen in Fig.~\ref{fig:coex}, the Voronoi construction again fails to capture non-triangular tiles, so the comparison focuses on SANN and mSANN. In the QC12-hexagonal coexistence, both variants correctly recover squares, triangles, and pentagons (the latter corresponding to defects). SANN, however, undercounts in square-rich regions, often connecting diagonals across squares and thereby reducing the apparent number of squares. By contrast, mSANN captures both the missing squares and most pentagons, though it occasionally merges adjacent triangles into quadrilaterals in hexagonal domains.

The QC12-square coexistence shows similar behaviour. Here mSANN more reliably identifies the defects, including two shapes of defects corresponding to irregular hexagons \cite{ulugol2025defects}, which SANN misses in the majority of cases. Voronoi remains highly effective in pure hexagonal domains but offers no advantage in the mixed regions.

Overall, these tests confirm the complementary strengths of the algorithms. Voronoi is best suited to purely hexagonal environments but cannot describe non-triangular tiles. SANN and mSANN are both capable of analyzing coexisting phases, with mSANN providing more consistent identification in low-CN regions and superior detection of defect structures.

\subsection{Performance analysis}

\begin{figure}
\includegraphics[width=\columnwidth]{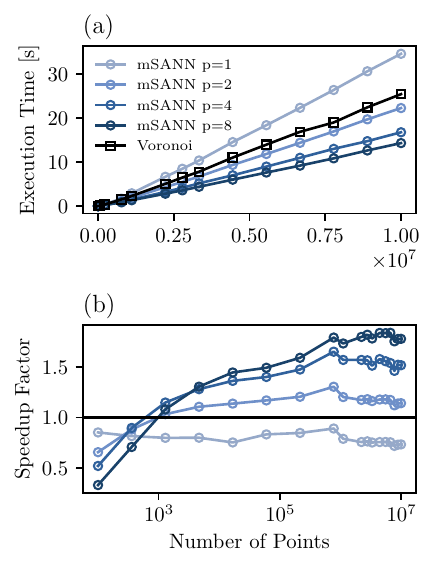}
\caption{Benchmark of our implementation of 2D mSANN versus Triangle's implementation of Voronoi construction. $p$ is the number of cores used for the execution. Panel (a) shows the total execution time of the algorithms as a function of points in the configuration. Panel (b) shows the speedup factor, defined as ratio of the execution times of the Voronoi algorithm and  the mSANN variants.}
\label{fig:bench}
\end{figure}

Low computational cost is a key requirement for a ``good'' neighbor-identification algorithm. To assess this property, we implemented the two-dimensional extension of SANN in Python and optimised it for speed. The computational bottlenecks are pair-distance evaluation and sorting, which scale as $\mathcal{O}(N^2)$ in naive implementations. To reduce this cost, we employ a $k$-d tree, which combines proximity search and sorting in $\mathcal{O}(N \log N)$. In addition, the algorithm is parallelized and just-in-time compiled using Numba, providing significant acceleration.

For benchmarking, we compared our implementation to Jonathan Richard Shewchuk's \texttt{Triangle} library \cite{shewchuk1996triangle}, which provides a highly optimized Voronoi construction algorithm. This package is, to our knowledge, the fastest available implementation of two-dimensional Voronoi tessellation and therefore serves as a suitable baseline. We generated systems containing between $10^2$ and $10^7$ points, distributed uniformly at random in the unit square. For each system size, we prepared $128$ independent realizations, and measured the average runtime and standard deviation. All calculations were performed on a machine running macOS 15.6 with an Apple M3 Max SoC and 64~GB of memory.

The benchmark results are shown in Fig.~\ref{fig:bench}. For systems smaller than $1000$ points, Voronoi outperforms both variants of SANN. For larger systems, however, the parallelized mSANN implementation consistently surpasses Voronoi, with speedups of up to $1.8\times$, while the sequential version remains the slowest across all system sizes. These results indicate that for small systems Voronoi remains the fastest option, but for simulations involving more than $10^3$ points, parallelized mSANN offers the best computational speed.

\section{Conclusion}

We have revisited the SANN algorithm and examined its extension to two dimensions. While the defining relation of SANN is linear in three dimensions, it becomes nonlinear in two dimensions, leading to increased computational cost. Here, we examined the  reformulated SANN introduced in Ref. \onlinecite{mugita2024simple} (SANNex) which approaches  the cutoff determination as an existence–uniqueness problem, and avoids solving the nonlinear equation directly while retaining the efficiency and general features of the original method.

We also analyzed the tendency of both SANN and Voronoi to overcount neighbors in low-coordination-number lattices, an effect amplified in two dimensions. To mitigate this, we introduced the inscribed circle (sphere) modification, a parameter-free modification based on the relation between inscribed and circumscribed radii of regular polygons (polyhedra), which we called mSANN. We show that the correction yields more intuitive and robust results in both 2D and 3D.

Tests on ordered lattices showed that mSANN is more likely to identify a number of neighbors equal to the coordination number in open lattices, such as honeycomb, square, simple cubic, diamond, and graphite. In contrast,  the original SANN remains mildly more robust in high-coordination environments such as hexagonal, FCC, and HCP lattices. Voronoi remained accurate in purely hexagonal regions but showed limitations elsewhere. In quasicrystalline tilings, mSANN typically directly yields the structure of the tiling, by avoiding the inclusion of diagonal bonds across squares seen in both Voronoi and SANN. For coexisting structures, mSANN provided consistent results across interfaces and defects. 

Finally, performance benchmarks demonstrated that for small systems ($N<10^3$) Voronoi remains faster, while for larger systems parallel mSANN consistently outperforms Voronoi by up to a factor of 1.8.  

In summary, the proposed mSANN algorithm offers a parameter-free, computationally efficient, and robust approach to nearest-neighbor identification in two and three dimensions, that is particularly appropriate for systems with low coordination numbers, coexistence, and/or defects.

\section*{Acknowledgements}
We would like to thank Geert H. A. Schulpen and Marjolein de Jager for fruitful discussions. A.U. and L.F. acknowledge funding from the XL research program with project number OCENW.GROOT.2019.071 which is financed by the Dutch Research Council (NWO).

\section*{Data Availability}
The data that support the findings of this study are openly available on Zenodo at \href{https://doi.org/10.5281/zenodo.18457313}{DOI:10.5281/zenodo.18457313}.  
The Python module implementing the original SANN algorithm and its modified variants is available at \href{https://github.com/alptug/SANNpy}{https://github.com/alptug/SANNpy}.

\appendix

\section{Relative Spread of Nearest Neighbors}\label{ap:spread}
To quantify the effect of breaking the equidistant-neighbor reference case underlying the inscribed-circle construction, we analyze the relative spread of nearest-neighbor distances,
$r_{i,m}/r_{i,1}$,
where $r_{i,1}$ and $r_{i,m}$ denote the distances to the closest and furthest nearest neighbors of particle $i$, respectively, as identified by mSANN.

This ratio provides a direct measure of how strongly a local environment deviates from the idealized equidistant-neighbor limit $r_{i,m}/r_{i,1}=1$. Values larger than unity indicate increasing heterogeneity in nearest-neighbor distances due to thermal fluctuations, disorder, or structural complexity.

\begin{figure*}
    \centering
    \includegraphics[width=0.9\linewidth]{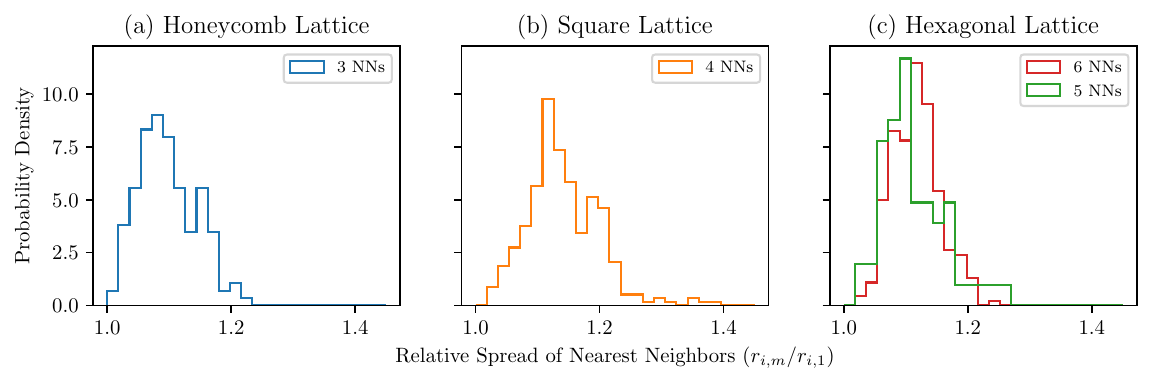}
    \caption{Distributions of the nearest-neighbor distance spread $r_{i,m}/r_{i,1}$ obtained with mSANN for self-assembled two-dimensional crystalline lattices, including honeycomb, square, and hexagonal structures. }
    \label{fig:spreadlat}
\end{figure*}

We first consider self-assembled crystalline lattices in two dimensions, including honeycomb, square, and hexagonal structures, perturbed by thermal fluctuations. The resulting distributions of $r_{i,m}/r_{i,1}$ are shown in Fig.~\ref{fig:spreadlat}, resolved by coordination number. As expected, thermal noise broadens the distributions away from unity. Importantly, all systems analyzed here deviate substantially from the equidistant-neighbor reference, yet mSANN consistently identifies the correct coordination numbers across the observed range of distance heterogeneity.

\begin{figure*}
    \centering
    \includegraphics[width=0.9\linewidth]{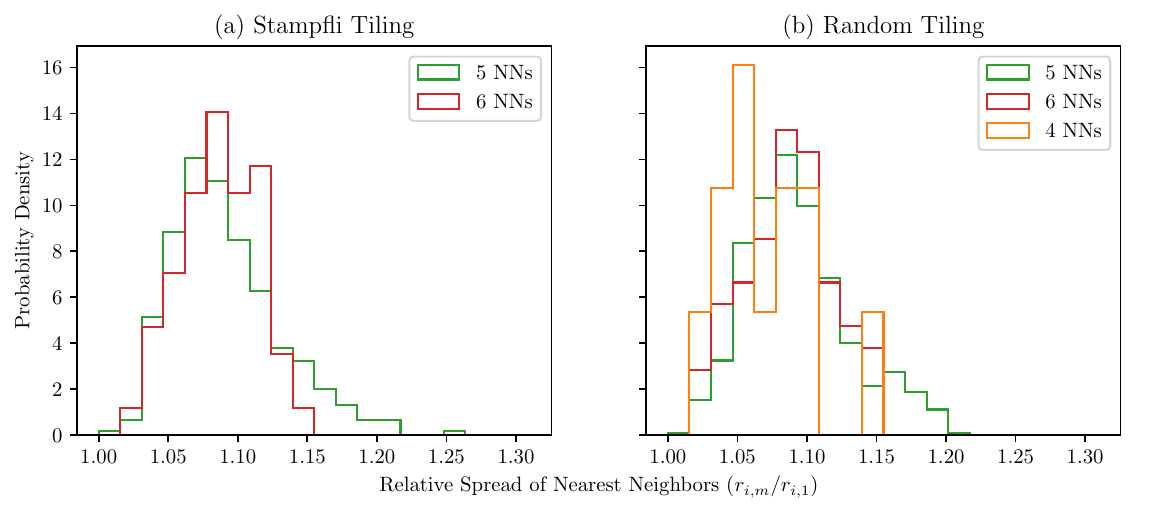}
    \caption{Distributions of the nearest-neighbor distance spread $r_{i,m}/r_{i,1}$ obtained with mSANN for 12-fold symmetric quasicrystal approximants. Results are shown separately for different coordination numbers present in the Stampfli tiling and in the randomized square–triangle tiling.}
    \label{fig:spreadqc}
\end{figure*}

We next analyze 12-fold symmetric quasicrystal approximants, including the Stampfli tiling and randomized square-triangle tilings, which exhibit intrinsic structural disorder even in the absence of thermal noise. The corresponding distributions of $r_{i,m}/r_{i,1}$ are shown in Fig.~\ref{fig:spreadqc}. These distributions appear to be a mixture of the square and the hexagonal lattice cases, reflecting the heterogeneous local environments characteristic of the square-triangle tiling. Nevertheless, mSANN remains robust across the empirically observed distance spreads and correctly resolves nearest neighbors for all coordination numbers present in these structures.

Taken together, these results demonstrate that mSANN does not rely on the equidistant-neighbor limit in practice. Instead, it remains robust across the range of nearest-neighbor distance heterogeneities naturally encountered in self-assembled crystalline and quasicrystalline systems. At the same time, these distributions highlight that the applicability domain of mSANN is governed by an interplay of coordination number, distance spread, and local geometry rather than by a single scalar threshold.

\section{An example for the broken equidistant-neighbor property}\label{ap:example}
Suppose that the closest $4$ points to particle $i$ are $r_{i,1}, r_{i,2}, r_{i,3}, r_{i,4}$ distance away and the relative distances are given by
\begin{equation*}
        r_{i,2} = 1.1125r_{i,1},\quad r_{i,3} = r_{i,4}=2.25r_{i,1}.
\end{equation*}
and
\begin{equation*}
k_m = \frac{1}{2}\big[1+\cos(\pi/m)\big].
\end{equation*}

Then, $f_i(r_{i,4};3)$ evaluates to
\begin{equation*}
\begin{aligned}
    f_i(r_{i,4};3) =& -\pi + \arccos\!\left(k_3\frac{r_{i,1}}{r_{i,4}}\right) + \arccos\!\left(k_3\frac{r_{i,2}}{r_{i,4}}\right)
    \\&+ \arccos\!\left(k_3\frac{r_{i,3}}{r_{i,4}}\right)
    \\=& -\pi + \arccos\!\left(1/3\right) + \arccos\!\left(0.3708\bar{3}\right)
    \\&+ \arccos\!\left(3/4\right)\\
    \approx &\, 0.003 > 0.
\end{aligned}
\end{equation*}
Therefore, the algorithm converges at $m=3$, although the third and the fourth neighbors are equidistant to particle $i$.
\bibliography{refs}

\end{document}